\documentclass[twocolumn,twocolappendix]{aastex63}

\received{\today}
\revised{\today}
\accepted{\today}
\submitjournal{ApJL}

\usepackage{amsmath, bm, amssymb}
\usepackage[abs]{overpic}
\usepackage{xcolor, varwidth, multirow, ctable, makecell}

\shorttitle{TeV flares of M87*}
\shortauthors{H. Hakobyan, B. Ripperda, A.A. Philippov}

\graphicspath{{./}{figures/}}

\definecolor{newred}{rgb}{0.86, 0.078, 0.24}
\newcommand{\NEW}[1]{{#1}}

\newcommand{\gammarad}{\gamma_{\rm rad}^{\rm syn}}

\begin{document}

\title{Radiative reconnection-powered TeV flares from the black hole magnetosphere in M87}

\correspondingauthor{H. Hakobyan}
\email{haykh.astro@gmail.com}

\author[0000-0001-8939-6862]{H. Hakobyan}
\affiliation{Computational Sciences Department, Princeton Plasma Physics Laboratory (PPPL), Princeton, NJ 08540, USA}
\affiliation{Physics Department \& Columbia Astrophysics Laboratory, Columbia University, 538 West 120th Street, New York, NY 10027, USA}
\author[0000-0002-7301-3908]{B. Ripperda}\altaffiliation[NASA Hubble Fellowship Program, Einstein Fellow]{}\affiliation{School of Natural Sciences, Institute for Advanced Study, 1 Einstein Drive, Princeton, NJ 08540, USA}\affiliation{Department of Astrophysical Sciences, Peyton Hall, Princeton University, Princeton, NJ 08544, USA}\affiliation{Center for Computational Astrophysics, Flatiron Institute, 162 Fifth Avenue, New York, NY 10010, USA}
\author[0000-0001-7801-0362]{A. A. Philippov}
\affil{Department of Physics, University of Maryland, College Park, MD 20742, USA}

\begin{abstract}

Active Galactic Nuclei in general, and the supermassive black hole in M87 in particular, show bright and rapid gamma-ray flares up to energies of 100 GeV and above. For M87, the flares show multiwavelength components, and the variability timescale is comparable to the dynamical time of the event horizon, suggesting that the emission may come from a compact region nearby the nucleus. However, the emission mechanism for these flares is not well understood. Recent high-resolution general-relativistic magnetohydrodynamic simulations show the occurrence of episodic magnetic reconnection events that can power flares nearby the black hole event horizon. In this work, we analyze the radiative properties of the reconnecting current layer under the extreme plasma conditions applicable to the black hole in M87 from the first principles. We show that abundant pair production is expected in the vicinity of the reconnection layer, to the extent that the produced secondary pair plasma dominates the reconnection dynamics. Using analytic estimates backed by two-dimensional particle-in-cell simulations we demonstrate that in the presence of strong synchrotron cooling, reconnection can produce a hard power-law distribution of pair plasma imprinted in the outgoing synchrotron (up to a few tens of MeV) and the inverse-Compton signal (up to TeV). We produce synthetic radiation spectra from our simulations, which can be directly compared with the results of future multiwavelength observations of M87* flares.



\end{abstract}


\section{\label{sec:level1}Introduction}

Active Galactic Nuclei (AGN) regularly show bright and rapid $\gamma$-ray flares (e.g., \citealt{Albert2007,Aharonian2007,Aharonian2009,Aleksic2014S}), up to very high energies ($> 100$ GeV). In particular, the black hole in the center of the M87 galaxy shows episodic multiwavelength flares, including TeV components (\citealt{Hess2006,Veritas2010,2011ApJ...743..177H,veritas2012,Magic2021}). The energy release of the highest energy flares, $\gtrsim 10^{41}$ erg/s (\citealt{Hess2012}), is non-negligible compared to the total jet power of $10^{42}-10^{44}$ erg/s (e.g., \citealt{Prieto2016}). The rapid variability timescales of the order of a few light-crossing times of the event horizon associated with the TeV emission from M87 imply that the emitting region is of the order of the Schwarzschild radius, within the uncertainty of an unknown beaming factor (\citealt{2011ApJ...743..177H}). Additionally, the existence of an X-ray emission counterpart  originating from nearby the nucleus for some of the events (\citealt{2011ApJ...743..177H, Hess2012, 2018A&A...612A.106S}) suggests that the origin of the flares is in the vicinity of the central black hole. 

The conversion of magnetic energy into kinetic energy and radiation through magnetic reconnection is an obvious and often proposed candidate to power flares from magnetized plasma nearby the event horizon. The exact emission mechanism that powers these TeV flares is however not well understood. 
To study whether magnetic reconnection can power the observed radiation, it is essential to know the geometry and strength of the magnetic field nearby the event horizon.
Event Horizon Telescope (EHT) observations of polarized radio emission \citep{EHTVII2021} show that the accretion disk of M87 is likely to be in a particular scenario uncovered by general relativistic magnetohydrodynamics (GRMHD) simulations: the magnetically arrested state (\citealt{narayan2003,Tchekhovskoy2011}). In this state, the magnetic flux on the black hole accumulates onto the event horizon until its pressure becomes dynamically important. When the magnetic pressure due to flux accumulation becomes comparable to the pressure of accreting gas, the flux is expelled through magnetic reconnection events in an equatorial current sheet that separates the northern and southern jet (\citealt{Ripperda2020,Ripperda2022}). 
Based on their unambiguous occurrence in global GRMHD simulations, these reconnection events have been conjectured to power high-energy flares from nearby the event horizon of the black hole \citep{Ripperda2020,Chashkina2021,Bransgrove2021,Ripperda2022}. 



GRMHD simulations, however, cannot predict the properties of the reconnection-powered emission. To understand the mechanisms that power the multiwavelength flaring emission, it is essential to rely on kinetic physics that describes particle acceleration from first principles. Moreover, the interaction between accelerated leptons and the emitted photons, which involves both classical (such as the radiation drag), and quantum effects (such as pair-production and Compton scattering), is unexplored in the regimes applicable for radiatively inefficient accretion flows in AGN.

Possible radiative mechanisms discussed in the literature to explain this emission often invoke curvature and inverse Compton (IC) radiation by leptons (electrons and positrons), accelerated to high energies in vacuum ``gap'' regions (\citealt{Levinson2011,Broderick2015,chen2019,Crinquand2020}). While being a promising source of TeV flares, gap models require an external trigger mechanism, which can substantially modify the properties of the background radiation field \citep{Kisaka22}. Considerable changes can lead to significant particle acceleration and flaring emission due to the development of large regions with an unscreened electric field, that appear because of the change in the optical depth to $\gamma\gamma$ pair-production. Magnetic reconnection, on the other hand, self-consistently occurs as a consequence of the magnetic flux accumulation near the black-hole horizon during accretion. However, since the reconnection is a microscopic process dissipating the magnetic field energy and accelerating particles on scales much smaller than the horizon scale, it is unclear whether it can power the observed TeV flares both in terms of the total luminosity and in terms of the maximum photon energy. 



\NEW{
In this Letter, we build a semi-analytic model for the flares in M87* and other AGN powered by the intermittent magnetic reconnection in the equatorial plane. We propose that short-duration high-luminosity TeV flares can indeed be produced via IC radiation powered by the upscattering of soft photons from the accretion flow by the electrons and positrons accelerated during magnetic reconnection. Our model further predicts a broadband spectrum ranging from radio to TeV. To quantify our findings, we conduct first-principles kinetic simulations of radiative reconnection in collisionless pair plasma, including dynamically important synchrotron radiation, and taking into account IC scattering in post-processing, for parameters applicable nearby the event horizon of the supermassive black hole in the center of M87. 
}


\section{M87* TeV flares: theory}
\label{sec:theory}
Any theoretical model aiming to explain the underlying physics behind the TeV emission is bound to incorporate three empirical characteristics of these flares: energetics, luminosity, and \NEW{fast rise and fall timescales}. First of all, in order to produce TeV emission it is necessary to at least have an efficient mechanism for particle energization able to accelerate leptons to energies $E_{e^\pm}\gtrsim 10^6m_e c^2$. \NEW{In the context of reconnection -- the process our discussion is centered upon -- the characteristic particle energy depends on the available magnetic energy per particle, which in turn depends on how much plasma is supplied to the reconnection region.} On top of that, the produced TeV flux, regardless of the underlying emission mechanism, has to be able to escape the magnetosphere. This fact puts additional upper limits for the optical depth to the Breit-Wheeler pair production process which can potentially inhibit the TeV photon flux, converting its energy into secondary $e^\pm$ pairs \cite[see, e.g.,][]{Levinson2011}. \NEW{Subsequently, given the relatively high luminosity of the flares (for M87 they reach $10^{39}\text{-}10^{40}$ erg/s) and the rapid rise/fall timescales (few days for M87), the model needs to invoke a very fast and efficient energy dissipation mechanism. As we discuss further in this section, radiative fast magnetic reconnection is an attractive solution in that regard, since it is able to dissipate the supplied magnetic energy rapidly and subsequently convert most of the liberated energy into the high-energy emission. }


Recent GRMHD simulations of magnetically arrested black hole accretion flow \citep{Ripperda2020,Chashkina2021,Ripperda2022} show clear indications of episodic magnetic reconnection events occurring close to the black hole event horizon.
During such a flux eruption, a large part of the accretion flow can be ejected, and a low-density highly magnetized ``magnetosphere'' forms near the black hole.
In GRMHD, this magnetospheric region upstream of the reconnection layer, i.e., at the jet base, consists of low-density plasma with a magnetization set by the density floor, indicating that pair plasma from the \NEW{broadened base of the} jet supplies the matter into the current sheet \citep{Ripperda2020,Ripperda2022}.


From first-principles kinetic simulations we know that even in the presence of strong radiation drag, magnetic reconnection renders itself as an effective process for particle acceleration \citep{2014ApJ...783L..21S, 2014PhRvL.113o5005G, 2016ApJ...816L...8W,2019ApJ...877...53H,2019MNRAS.482L..60W, 2020ApJ...899...52S}. As we demonstrate below, reconnection near the event horizon of M87$^{*}$ occurs in the radiative regime, when the synchrotron cooling timescale is comparable to the particle acceleration time. In this case, most of the dissipated magnetic field energy is radiated as synchrotron photons with a broad spectrum peaking around $\varepsilon_c= (3 \beta_{\rm rec}/\alpha_F) m_e c^2\approx 20$ MeV, where $\alpha_F$ is the fine-structure constant, and $\beta_{\rm rec}\approx 0.1$ is a dimensionless number characterizing the rate at which reconnection occurs in the collisionless regime. If the luminosity of the synchrotron emission is large, the plasma content of the current sheet will self-consistently be regulated by pair production fueled by these \NEW{synchrotron photons spanning from keV to GeV energies}. The important dimensionless parameter that determines the efficiency of pair acceleration in reconnection is \NEW{(twice)} the available magnetic energy per particle, otherwise referred to as the \emph{magnetization parameter}: $\sigma = B^2/(4\pi \rho_{e^\pm}c^2)$, where $B$ is the characteristic strength of the unreconnected magnetic field (i.e., the field upstream of the current sheet, in the jet), and $\rho_{e^\pm}$ is the density of pair-plasma. 

The characteristic optical depth for the photons interacting with the other synchrotron photons of average energy $\varepsilon_1$ and luminosity $L$ is: $\tau_{\gamma\gamma}\approx f_{\gamma\gamma}\sigma_T\dot{n} (w/c) w\approx 3f_{\gamma\gamma}\sigma_T L / (4\pi c w \varepsilon)$, where we assumed that $(4\pi w^3/3)\dot{n}\approx L / \varepsilon$, $w\sim 10~r_{g}$ is the typical length scale of the reconnecting region, and $f_{\gamma\gamma}\approx 0.25$ (set by the maximum $\gamma\gamma$ cross section; see, e.g., \citealt{1985quel.book.....A}). \NEW{To get a rough estimate on the fraction of the radiated photons being converted back into pairs, we evaluate the fiducial optical depth, assuming $L\approx f_{\rm rec}L_{\rm jet}$ (fraction of the jet power), and $\varepsilon\approx m_e c^2$: $\tau_{\gamma\gamma}^0\approx 3f_{\rm rec}f_{\gamma\gamma}\sigma_T L/(4\pi m_e c^3 w)\sim 10^{-4}\ll 1$. As it becomes clear in sec.~\ref{sec:rad_drag}, most of the photons that participate in the pair-production process also happen to be the ones that carry most of the dissipated power. This suggests that the fiducial value for the optical depth is a good proxy for the actual population-averaged optical depth, $\tau_{\gamma\gamma}^0\approx \tau_{\gamma\gamma}$.} \NEW{The amount of produced pairs (over the whole magnetosphere) can then be estimated as $\dot{N}_{e^\pm} \approx (1/2)\sigma_T L^2/(w c\varepsilon_c^2)$ (for a detailed derivation see the appendix~\ref{app:pp}), where we implicitly assumed that most of the synchrotron energy is carried by the photons with energies around a few-to-ten MeV, i.e., $\varepsilon_c$.}\footnote{\NEW{The justification of this ansatz will become clear later in the section~\ref{sec:rad_drag}, where we show that the peak of the emission is slightly above MeV. Even for a relatively broad distribution this formula is a reasonable approximation for the total pair-production efficiency, as long as the peak is around $m_e c^2$ and the optical depth is small (see, e.g., \citealt{Hakobyan2019}). }}

 
 To obtain the multiplicity of the produced plasma with respect to the Goldreich-Julian number density, we can compare $\dot{N}_{e^\pm}$ with the Goldreich-Julian number flux, $\dot{N}_{\rm GJ}\approx \left(c L_{\rm jet}\right)^{1/2}/|e|$:\footnote{\NEW{To estimate the Goldreich-Julian number flux, we take $\dot{N}_{\rm GJ}\approx I_{\rm GJ}/|e|$, where $I_{\rm GJ}$ is the magnetospheric current induced by the rotation of the black hole in the jet region. From Maxwell's equation for the toroidal component of the magnetic field, $(4\pi/c)I_{\rm GJ}\approx 2\pi w B_\phi$, where $B_\phi$ also enters into the expression for the Poynting flux: $S_P\approx (c/4\pi) B_\phi^2$. Connecting the Poynting flux with the jet power, $L_{\rm jet}\approx \pi w^2 S_P$, we then arrive at $I_{\rm GJ}\approx \sqrt{cL_{\rm jet}}$.}}

\begin{figure*}[htb]
    \centering
    \includegraphics[width=\textwidth,trim={0 10 0 10},clip]{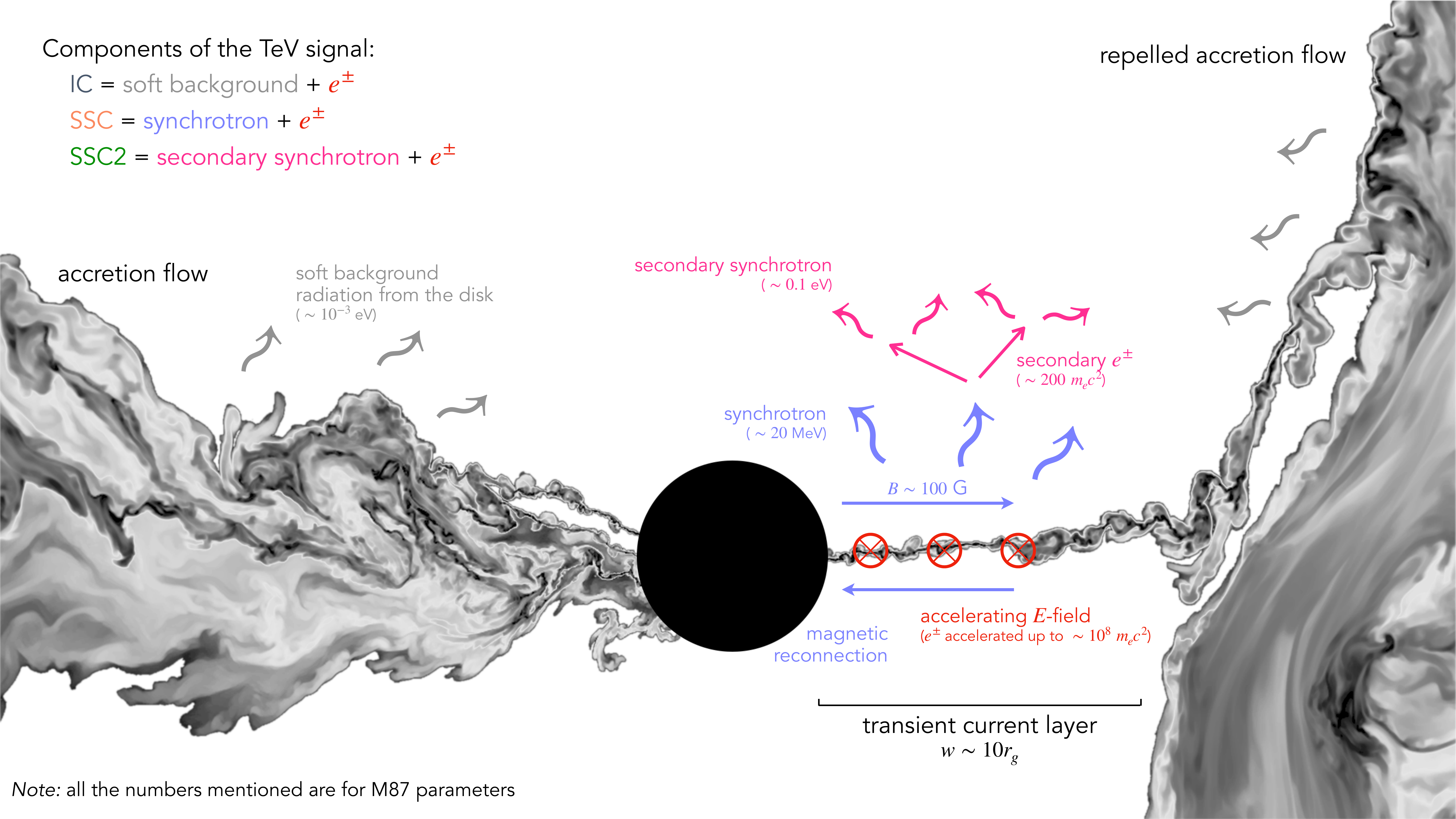}
    \caption{Schematic illustration of the black hole accretion flow during a reconnection-mediated flare. During the flare, the accretion flow is repelled outwards by the magnetic field to about $\sim 10r_g$. A transient current layer is formed in the equatorial plane, where magnetic reconnection takes place. Leptons are accelerated to large energies in the current sheet, producing high-energy synchrotron emission (shown in blue). Synchrotron photons interact with each other producing $e^\pm$-pairs, which are fed back to the current layer. These secondary pairs in turn radiate much lower-energy synchrotron radiation (shown in pink). The TeV signal is produced by the upscattering of soft photons of different origins by the accelerated pairs in the current layer. All the colors used in this figure correspond to the ones in Figure~\ref{fig:m87_predict}. The snapshot of plasma-$\beta$ is taken from the high-resolution GRMHD simulation of the magnetically arrested accretion state by \cite{Ripperda2022}.}
    \label{fig:sketch}
\end{figure*}

\begin{widetext}
\begin{equation}
\label{eq:multiplicity}
\frac{\dot{N}_{e^\pm}}{\dot{N}_{\rm GJ}}\sim 10^6\left(\frac{f_{\rm rec}}{0.1}\right)^2\left(\frac{L_{\rm jet}}{10^{43}~\text{erg/s}}\right)^{3/2}\left(\frac{M_{\rm BH}}{6\cdot 10^9~M_\odot}\right)^{-1}\left(\frac{w/r_g}{10}\right)^{-1},
\end{equation}
\end{widetext}
where we assume that a fraction of the Poynting flux carried by jet, $L_{\rm rec}\approx f_{\rm rec}L_{\rm jet}$, is dissipated during the reconnection event and is radiated away, and $M_{\rm BH}$ and $r_g = GM_{\rm BH}/c^2$ are the mass and the gravitational radius of the black hole, normalized to fiducial values for M87. This estimate is well justified in the regime where the effective optical depth to pair production is small, $\tau_{\gamma\gamma}\sim 10^{-4}$, and both high and low energy photons participating in pair production are produced by the same dissipation mechanism. We demonstrate a more rigorous estimation for the $f_{\rm rec}$ parameter in sec.~\ref{sec:syncLuminosity}.
Based on this estimate, one can clearly see that the dynamics of the reconnecting current sheet is controlled by the density of the pair plasma $n_{e^\pm}\approx 10^6 n_{\rm GJ}$ produced \emph{in-situ}, which is much larger compared to the density produced by gaps or pair ``drizzles'' in the jet \citep{Moscibrodzka2011, Crinquand2020, chen2019, 2021ApJ...907...73W, 2021MNRAS.507.4864Y}. This yields a value for the plasma magnetization:

\begin{equation}
    \label{eq:sigma_th}
    \sigma \approx \frac{\omega_B}{2\Omega_{\rm BH}}\frac{\dot{n}_{\rm GJ}}{\dot{n}_{e^\pm}}\sim 5\cdot10^7\left(\frac{B}{100~\text{G}}\right) \left(\frac{L_{\rm jet}}{10^{43}~\text{erg/s}}\right)^{-3/2},
\end{equation}
where $\omega_B=|e|B/m_e c$, and $\Omega_{\rm BH}\approx \left(2r_{g}/c\right)^{-1}$. In the situation when synchrotron cooling 
is dynamically important, pairs are able to accelerate to energies $E_{e^\pm}\approx \sigma m_e c^2$, forming a hard power-law energy distribution $f_{e^\pm}(\gamma)\propto \gamma^{-1}\text{-}\gamma^{-1.5}$, with $\gamma$ being the Lorentz factor of the pairs \NEW{(we demonstrate this directly from simulations in sec.~\ref{sec:simulations}; also see, e.g., \citealt{2019ApJ...877...53H}, and \citealt{chernoglazovinprep})}. A complete schematic illustration of the current sheet with all the sources for photons is shown in Figure~\ref{fig:sketch}. In the next sections we discuss the main radiation mechanisms and their energetics during the transient reconnection event.


\subsection{The role of the radiation drag}
\label{sec:rad_drag}

To quantify the effects of cooling and its dynamical importance with respect to the acceleration by reconnection, it is helpful to define a dimensionless parameter, $\gamma_{\rm rad}$, as $E_{\rm rec} |e| \equiv 2\sigma_T U \gamma_{\rm rad}^2$, where the left-hand side is the accelerating force experienced by particles in the reconnection sites, while the right-hand side corresponds to a radiation drag force either due to synchrotron or IC (where we assume the Thomson regime and $\sigma_T$ is the Thomson cross section) radiation. Here, $E_{\rm rec}$ is the electric field strength in the so-called x-point regions of the reconnecting current sheet, where the bulk of particle energization and energy dissipation takes place. Its value is controlled by the strength of the reconnecting magnetic field, and the rate at which the magnetic energy is dissipated (\emph{reconnection rate}); it can be written as $E_{\rm rec}\approx \beta_{\rm rec}B$. We take $\beta_{\rm rec}=0.1$, which is characteristic for collisionless \NEW{relativistic} reconnection. 
The value of $U$ in the definition of $\gamma_{\rm rad}$ can either be interpreted as the energy density of the magnetic field, $U=U_B\equiv B^2/8\pi$, or as the energy density of the background (seed) photon field, depending on whether we are considering synchrotron or IC cooling. When the forces are balanced, particles with energies $\gamma \gg \gamma_{\rm rad}$ will cool down faster (either via synchrotron or IC) than they are able to accelerate, whereas cooling will be negligible \NEW{on acceleration timescales} for particles with $\gamma \ll \gamma_{\rm rad}$. Thus, comparing $\gamma_{\rm rad}$ with the magnetization that powers the acceleration, $\sigma$, offers an important insight for understanding the cooling efficiency in the context of acceleration by reconnection.

For synchrotron radiation, which for M87 is by far the strongest cooling mechanism, the parameter can be estimated as $\gammarad = (4\pi\beta_{\rm rec}|e|/( \sigma_T B))^{1/2}\sim 4\cdot 10^6\left(B/100~\text{G}\right)^{-1/2}$. One can already draw two important conclusions from these estimates. First, since $\sigma\gtrsim \gammarad$, the reconnection proceeds in the radiative regime where most of the dissipated magnetic energy is quickly converted into synchrotron radiation on timescales much shorter than the dynamical time of the system. This suggests that the energy density of the synchrotron photons in the steady state is comparable to the dissipated magnetic energy density, \NEW{$U_{\rm rad}^{\rm syn}\approx \beta_{\rm rec} U_B$}. Note that the synchrotron cooling is negligible in x-points, where the magnetic field vanishes, meaning that the fast acceleration process is uninhibited by synchrotron radiation; the highest energy particles will still accelerate to $E_{e^\pm}\approx \sigma m_e c^2$. The peak of the synchrotron emission will then be set by the burnoff limit: $\varepsilon_c\approx \hbar \omega_B\left(\gammarad\right)^2\sim 20$ MeV \citep{2014ApJ...780....3U}.\footnote{Notice that in this particular regime the energy of the synchrotron peak is independent of any of the physical quantities and is essentially a universal constant: $\varepsilon_c\approx (3\beta_{\rm rec}/\alpha_F) m_e c^2$, where $\alpha_F\equiv e^2/(\hbar c)\approx 1/137$, and $\beta_{\rm rec}\equiv 0.1$ is the fiducial reconnection rate.}

\subsection{Radiation from secondary pairs}
\label{sec:sec_sync}

The optical depth for synchrotron photons produced in reconnection is small $\tau_{\gamma\gamma}\ll 1$ for M87 \citep{Ripperda2022}, yielding a very non-local pair production in the whole region of size $w$ in the black hole magnetosphere. These pairs, produced by photon-photon collisions, have arbitrary pitch angles to the local magnetic field and thus radiate rapidly giving rise to a distinct synchrotron spectral component different from the synchrotron continuum from reconnection. As the main fuel for these secondary pairs are the synchrotron photons from reconnection, their characteristic Lorentz factors at the time of birth can be estimated as $\gamma_{\rm sec}\approx (\text{few}\cdot\varepsilon_c) / (m_e c^2)\sim 200$. The synchrotron cooling timescale for the secondary pairs, $t_{\rm sec}$, compared to the light crossing time, is $t_{\rm sec}c/w\approx  (\gamma_{\rm sec}l_B)^{-1}\sim 10^{-3}$, where $l_B=\sigma_T U_B w/(m_e c^2)\sim 1$ is the magnetic compactness parameter \citep{Beloborodov2017}, which means that a significant fraction of the energy contained in these pairs is radiated away inside the magnetosphere. The synchrotron peak energy associated with these pairs is $\varepsilon_{\rm sec}\approx \hbar\omega_B\gamma_{\rm sec}^2\sim 0.01\text{-}0.1$ eV, with the energy density of resulting photons being controlled by the optical depth $U_{\rm rad}^{\rm sec}\approx U_{\rm rad}^{\rm syn}\tau_{\gamma\gamma}\sim 0.006$ erg$/$cm${}^3$, and the luminosity being close to $L_{\rm rad}^{\rm sec}\approx L_{\rm rec} \tau_{\gamma\gamma}\sim 10^{38}$ erg/s.

\subsection{Generation of the ultra-high energy signal}
\label{sec:TEVsignal}

Pairs accelerated by reconnection to energies $E_{e^\pm}\gtrsim 10^6\text{-}10^7 m_e c^2$ can Compton-upscatter lower energy photons to TeV energies (the upper limit will be determined by the maximum pair energy). There will be several variations of this TeV signal. First of all, the reconnection region is filled with a soft background photon field from the accretion disk with average energies around $\varepsilon_{\rm soft}\approx 0.001$ eV ($300$ GHz) and a characteristic energy density $U_{\rm rad}^{\rm soft}\approx 0.01~\text{erg}/\text{cm}^3$ \citep{Broderick2015,MWL2021}. On the other hand, the region is also filled with synchrotron photons both from the pairs accelerated in reconnection and from the secondary pairs produced in the magnetosphere (as discussed in the previous section). Primary synchrotron photons from reconnection possess a wide range of energies up to a few $\varepsilon_c\approx 10\text{-}100$ MeV. The energy density of these photons is comparable to magnetic field energy density $U_{\rm rad}^{\rm syn}\approx \beta_{\rm rec} U_B\sim 40~\text{erg}/\text{cm}^3$, whereas the energy density of the secondary synchrotron photons is the aforementioned $U_{\rm rad}^{\rm sec} \sim 0.006$ erg$/$cm${}^3$. 

A TeV signal can be produced by upscattering of either of the photon components populating the reconnection region by the highest energy pairs with $\gamma\gtrsim 10^6$: the soft background (we will refer to this channel as IC), the primary synchrotron photons (synchrotron self-Compton, SSC), and the secondary synchrotron photons (secondary synchrotron self-Compton, SSC2). In all of these cases most of the power will be focused in the narrow frequency range corresponding to the highest energy pairs, i.e., above a few hundred GeV~\citep{1970RvMP...42..237B}. 

We will assume that the reconnection produces a power-law distribution of pairs with a cutoff, $f_{e^\pm}(\gamma)\propto \gamma^{-p} e^{-\gamma/\gamma_c}$, where the cutoff is $\gamma_c\approx \sigma$, and $1<p<1.5$~\citep{chernoglazovinprep}, \NEW{an assumption that we later verify using kinetic simulations}. Further in this section, we will be comparing the Compton luminosities peaking in TeV (IC, SSC, and SSC2) with the synchrotron luminosity which peaks at few tens of MeV. The power density of this emission is equal to $P_{\rm syn}\approx 2\sigma_T c n_{e^\pm}\langle\gamma^2 \tilde{B}_\perp^2/(8\pi)\rangle$, where the averaging is done over the particle distribution $\langle x\rangle \equiv \int d\gamma f_{e^\pm}(\gamma)x$, and $\tilde{B}_\perp$ is the effective magnetic field strength perpendicular to the motion of the particle (see eq. \eqref{eq:sync-force} for the full definition). As demonstrated in our simulations \NEW{in sec.~\ref{sec:simulations},} 
we can accurately approximate $\langle\gamma^2 \tilde{B}_\perp^2/(8\pi)\rangle\approx \chi^2\left(\gammarad\right)^2 U_B$ (as long as the cooling is strong), with $U_B$ being the magnetic energy density in the unreconnected upstream, and $\chi\approx 1$ being a dimensionless parameter of order unity. 
Using that, one finds $P_{\rm syn}\approx 2\sigma_T c n_{e^\pm}\chi^2\left(\gammarad\right)^2 U_B$.

\subsubsection{Total synchrotron luminosity}
\label{sec:syncLuminosity}

In the previous section, we assumed that the radiated synchrotron power during the reconnection is a non-negligible fraction of the jet power ($L_{\rm syn}\approx L_{\rm rec}\approx f_{\rm rec}L_{\rm jet}$, where $f_{\rm rec}\sim 0.1$). To obtain this estimate more rigorously let us multiply the emitted synchrotron power density by the volume to get $L_{\rm syn}=P_{\rm syn} \alpha \pi w^2 \delta_{\rm cs}/(2\pi)$, where $\alpha\sim 1$ is the azimuthal extent of the region undergoing reconnection, $w$ and $\delta_{\rm cs}$ are the radial extent and the thickness of the current sheet. As we demonstrate from the simulations in \NEW{sec.~\ref{sec:simulations}}, the characteristic temperature (or the mean energy) of particles in the current sheet is dictated by the synchrotron burnoff limit: $3T_{e^\pm}= \left(\kappa\gammarad\right) m_e c^2$, where $\kappa$ is a dimensionless parameter defined as follows: $\langle\gamma\rangle=\kappa\gammarad$. Given that, we can eliminate the number density, $n_{e^\pm}$, \NEW{using} the pressure balance condition: $n_{e^\pm}\approx U_B/T_{e^\pm}$. 
The width of the current sheet in collisionless reconnection is set by the characteristic Larmor radius of particles \citep[e.g.,][]{2014ApJ...780....3U}: $\delta_{\rm cs}\approx \langle\gamma\rangle m_e c^2/(|e|B)$. The total radiated synchrotron power then reads: 
\begin{equation}
\label{eq:sync_luminosity}
\begin{split}
    L_{\rm syn}\approx \frac{9\alpha}{2\pi} \beta_{\rm rec} 
        \underbrace{
            \frac{\langle\gamma^2 \tilde{B}_\perp^2\rangle}{\left(\gammarad\right)^2 B_0^2}
        }_{\chi^2}
        L_{\rm BZ}
        \approx 
            \underbrace{
            1.4\cdot 
            \alpha
            \chi^2
            \beta_{\rm rec}
            }_{\sim 0.1} L_{\rm BZ},
\end{split}
\end{equation}
where $L_{\rm BZ}\equiv B_0^2 r_g^2 c / 24$ is the \cite{BZ1977} luminosity for a maximally spinning black hole, \NEW{and we further assumed that the magnetic field falls with the distance from the horizon: $B=B_0(r/r_g)^{-1}$}. Notice that the dependence on $w$ has disappeared, due to the fact that the magnetic field strength decreases linearly with distance, and so does the effective value of $\gammarad$. Provided that the jet power is close to the BZ value $L_{\rm jet}\approx L_{\rm BZ}$, we can state that radiatively efficient reconnection has a potential of emitting about $10\%$ of the jet power, i.e., indeed, $f_{\rm rec}\sim 0.1$.

\vspace{1cm}
\NEW{As discussed earlier in this section, the bulk of the radiated power, $L_{\rm syn}$, will be carried by the photons with energies close to the burnoff limit, $\varepsilon_c\sim 20$ MeV. The very-high-energy (TeV) signal in our model is produced via Compton upscattering of lower-energy photons by the accelerated pairs, and will thus contain only a small fraction of the total synchrotron power. To estimate the power in the TeV component of the observed emission, in the following sections \ref{sec:soft_photons_ic}, \ref{sec:ssc2}, and \ref{sec:ssc} we investigate three different channels, which essentially consist of three populations of photons that can be upscattered to produce the TeV emission. }

\subsubsection{Upscattering of soft photons from the disk}
\label{sec:soft_photons_ic}
\NEW{The IC scattering of soft disk photons occurs in the Thomson regime}, as the typical energies of photons in the lepton rest-frame, are small, i.e., $\Gamma_{\varepsilon} = 4\gamma\varepsilon/(m_e c^2)\lesssim 1$, with $\varepsilon$ being the energies of soft background photons in the lab frame before the scattering, $\varepsilon/(m_e c^2)\approx 10^{-9}$, and $\gamma\approx\sigma\sim 10^7$ being the characteristic Lorentz factors of the leptons. The power per unit volume for the TeV signal can, thus, be found to be \citep[see, e.g.,][]{1979rpa..book.....R} $P_{\rm IC} \approx (4/3)\sigma_T c n_{e^\pm} \langle\gamma^2\rangle U_{\rm rad}^{\rm soft}$. For $\gamma_c\gg 1$ and $1<p<1.5$ we can approximate the average lepton energy to be the cutoff energy, i.e., $\langle\gamma^2\rangle \approx \gamma_c^2$. We may then compare this power density to the total synchrotron power density to obtain $P_{\rm IC} / P_{\rm syn}\sim 10^{-3}~\mathcal{U}_{0.01}^{(\rm soft)}\mathcal{B}_{100}^{-2}\mathcal{F}_{0.1}^{-2}$, where we assumed that most of the IC power is generated by particles close to the cutoff energy, $\gamma_c \sim \sigma$.\footnote{For the sake of brevity we employ the following notation: $\mathcal{U}_{0.01}\equiv U_{\rm rad}^{\rm soft}/(0.01~\text{erg}~\text{cm}^{-3})$, $\mathcal{B}_{100}\equiv B/(100~\text{G})$, $\mathcal{F}_{0.1}\equiv (\gamma_{\rm rad}^{\rm syn}/\gamma_c)/0.1$.}

\subsubsection{Upscattering of synchrotron photons from secondary pairs}
\label{sec:ssc2}

Photons from the fast cooling of the secondary pairs have energies $\varepsilon_{\rm sec}\approx 0.01\text{-}0.1$ eV. Collisions with the most energetic particles will occur in the regime intermediate between the Thompson and Klein-Nishina scattering approximations, where $\Gamma_{\varepsilon}\approx 0.1\text{-}1$. As an upper bound for the power we can use the same Thomson scattering relation as before, $P_{\rm SSC2}\approx (4/3)\sigma_T c n_{e^\pm} \langle\gamma^2\rangle U_{\rm rad}^{\rm sec}$. More realistically, one may use an approximation for the $\Gamma_{\varepsilon}\approx 1$ regime: $P_{\rm SSC2}\approx (1/48)\sigma_T c n_{e^\pm} (m_e c^2)^2 U_{\rm rad}^{\rm sec} / \varepsilon_{\rm sec}^2$; see eq.~\eqref{eq:lowpeak_ssc} in the appendix \eqref{app:ssc}. From these two estimates we find that $P_{\rm SSC2}/P_{\rm sync}\sim 10^{-6}\text{-}10^{-3}$.

\subsubsection{Upscattering of reconnection-produced synchrotron photons}
\label{sec:ssc}

The synchrotron photon field contains a much larger energy density than the soft background photons, $U_{\rm rad}^{\rm syn}\sim \beta_{\rm rec} U_B \gg U_{\rm rad}^{\rm soft}$. However, most of this energy is contained in photons around a few to a few tens of MeV, $\varepsilon_c$. In this case most of the SSC power is generated by the upscattering of low energy photons with energies $\varepsilon_s\lesssim 1~\text{eV}$ by the highest energy particles, so that $\Gamma_{\varepsilon}\approx 1$.\footnote{\NEW{Here we assume that the distribution of angles between the momenta of pairs and those of the photons is uniform. This assumption does not have a big impact on the amplitude of the outgoing very-high-energy flux, as we argue in the appendix~\ref{app:beamed-ssc}.}} In this regime neither the Thomson ($\Gamma_\varepsilon\ll 1$) nor the Klein-Nishina ($\Gamma_\varepsilon\gg 1$) approximation is valid. For a power-law distribution of particles one can estimate the power to be close to $P_{\rm SSC}\approx (3/4)\sigma_T c n_{e^\pm} \gamma_c^{(p+1)/2} (m_e c^2)U_{\rm rad}^{\rm syn}/\langle\varepsilon_s\rangle$, (see eq.~\eqref{eq:highpeak_ssc} in appendix \ref{app:ssc}), where $\langle\varepsilon_s\rangle\approx \varepsilon_c$ is the average energy of synchrotron photons. The value of this power is very sensitive to the shape of the power-law, $p$, as the latter determines the relative amount of photons with energies $\varepsilon_s\lesssim 1~\text{eV}$. From this approximation we deduce that the SSC power compared to the synchrotron power for values of the power law $1<p<1.5$ is $P_{\rm SSC}/P_{\rm sync}\sim 10^{-9}\text{-}10^{-7}$, which is clearly subdominant compared to the IC signal.

\subsection{Escape of the high-energy signal}

Pair production may greatly inhibit the observed luminosity of both the synchrotron MeV and the inverse-Compton TeV signal if the optical depth for these photons is large. As we estimated above, the optical depth for MeV photons interacting with each other is small, $\tau_{\gamma\gamma}\sim 10^{-4}$. Outgoing TeV photons, however, will interact with $\sim 1$ eV photons from the soft background. We may estimate the optical depth to that process similarly: $\tau_{\rm TeV}\sim 0.1\sigma_T w U_{\rm rad}^{\rm eV}/(1~\text{eV})$, where $U_{\rm rad}^{\rm eV}$ is the energy density of background photons with energies around $1$ eV. Given the total energy density of the soft radiation, $U_{\rm rad}^{\rm soft}$, as well as the slope of the distribution beyond $\varepsilon_{\rm soft}\approx 10^{-3}$ eV, $\varepsilon dn/d\varepsilon \propto \varepsilon^{-1.2}$ \citep{Broderick2015}, we can take $U_{\rm rad}^{\rm eV} \approx U_{\rm rad}^{\rm soft} \left(\varepsilon_{\rm soft}/1~\text{eV}\right)^{0.2}$. We then find an upper limit for the optical depth: $\tau_{\rm TeV}\approx 1$, meaning that a large fraction of the TeV signal escapes the inner few $r_g$.

\begin{figure*}[htb]
    \centering
    \includegraphics[width=\textwidth]{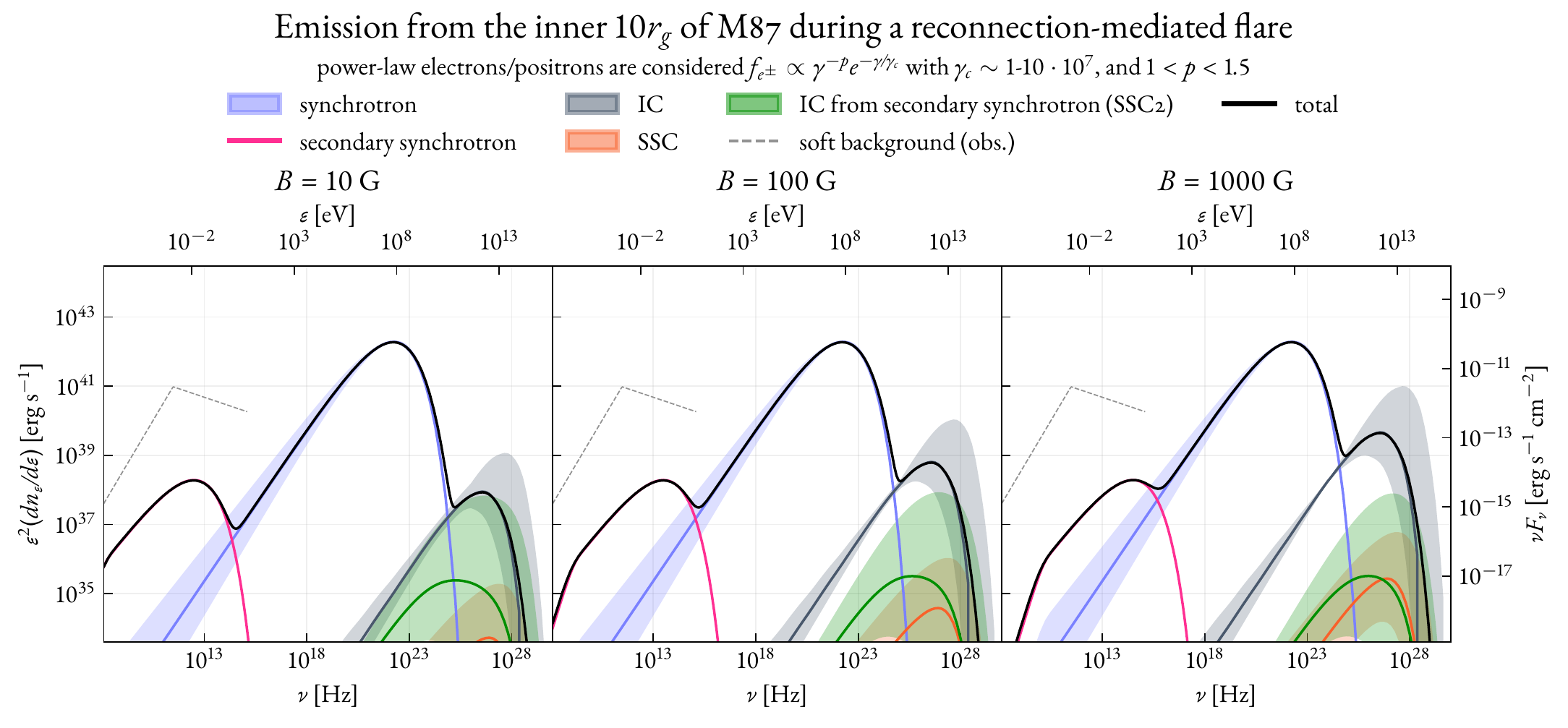}
    \caption{Predicted photon spectra from the reconnecting current sheet during the flaring state of M87. The blue line indicates the main synchrotron component, produced by particles having a power-law distribution with an index $p$ and cutoff $\gamma_c$. The pink line is the synchrotron emission from the secondary pairs produced outside the current sheet. The dashed gray line corresponds to the soft quiescent background emission. Compton-scattered photons dominate at energies $10\text{-}100$ GeV and above; the dark gray line corresponds to the upscattering of soft background photons (IC), the \NEW{orange line corresponds} to the synchrotron self-Compton (SSC) component, and the green line is the spectrum of upscattered secondary synchrotron photons (SSC2). In our calculations, we fix the luminosity of the synchrotron emission, and vary the effective magnetic field, $10<B~[\text{G}]<10^3$, the power-law index, $1<p<1.5$, and the cutoff energy of the accelerated pairs, $10^7<\gamma_c<10^8$.}
    \label{fig:m87_predict}
\end{figure*}

\vspace{0.5cm}

Our predictions are shown in Figure~\ref{fig:m87_predict}, where we plot the emerging spectra from all the emission components of the reconnecting sheet. In this plot we fix the main (synchrotron) luminosity to $10^{43}$ erg/s, while other components scale appropriately. \NEW{The synchrotron component from the secondary pairs (magenta line) is roughly estimated by taking a power-law distribution of pairs with a cutoff around $\gamma_{\rm sec}$ (see sec.~\ref{sec:sec_sync}). } We vary a few parameters of the problem: the effective magnetic field strength, $B$, in the region where most of the synchrotron emission takes place,\footnote{This value may be different from the average magnetic field in the black hole magnetosphere in the entire $<10 r_g$ region, since synchrotron cooling mainly takes place in the current sheet, where the magnetic field may be compressed or weakened.} the power-law index of the accelerated pairs, $p$, and their cutoff energy, $\gamma_c$. For comparison, we overplot the spectrum of soft photons from the disk at energies $\lesssim 10^{16}$ Hz from~\cite{Broderick2015}.

The main takeaway of these estimates is that the outgoing TeV signal is comprised of the IC scattering of soft background photons from the disk, synchrotron photons produced in-situ by the reconnecting sheet, and from the cooling of secondary pairs outside the reconnecting sheet. Our estimates show that the Compton scattering of soft background photons (IC component) contributes most to the very high energy signal, with the total TeV luminosity up to $\sim 0.01\text{-}0.1\%$ of the jet power, or about $10^{39}\text{-}10^{41}~\text{erg/s}$. Synchrotron radiation from secondary pairs appears as a distinct component at energies $\sim 0.01\text{-}0.1$ eV, with a luminosity up to $10^{38}\text{-}10^{39}~\text{erg/s}$.

\section{Simulations}
\label{sec:simulations}

\NEW{Our reconnection-powered flaring model described in the previous section relied on multiple quantitative assumptions and relations, the validity of which can only be tested by first-principles simulations.} In this section we describe the results from 2D particle-in-cell (PIC) simulations of relativistic radiative reconnection\NEW{, performed to verify the validity of these assumptions}. The simulations are performed using \texttt{Tristan v2} -- a multi-species radiative PIC code developed by~\cite{tristanv2}.
\vspace{1cm}

\subsection{Setup}

All of our simulations are initialized with a cold ($T_e\ll m_e c^2$) uniform background pair-plasma of density $n_0$, and a hot, thin, overdense layer of width $\Delta$ along the $x$-coordinate in the middle of the box (at $y=0$). The magnetic field is initialized as $\bm{B}=B_0\hat{\bm{x}}\tanh{\left(y/\Delta\right)}$ (where $B_0$ is the magnetic field strength in the far upstream), and $\bm{E}=0$ everywhere. The width, the temperature, and the density of the layer, as well as the in- and out-of-plane velocities of the particles in the layer are chosen in such a way that the setup is initially in the Harris equilibrium \citep{Harris1962}. Near the x-boundaries of the box we impose absorbing boundary conditions that damp the fields and absorb outgoing particles \citep{2016MNRAS.457.2401C}. Along the $x$-axis at a fixed distance from the layer we inject leptons to replenish the upstream plasma. These boundary conditions allow us to run our simulations for many light-crossing times of the box, ensuring the results are insensitive to the initial conditions. 

In all of our simulations, the value of the magnetization in the upstream is fixed: $\sigma = B_0^2/\left(4\pi n_0 m_e c^2\right)=10^3$. The size of the box is fixed at a value of $160\times 80$ in units of $\sigma\rho_0$, where $\rho_0=m_e c^2 / \left(|e| B_0\right)$ is the Larmor radius of fiducial particles with a velocity $\gamma\beta = 1$. The value $\sigma\rho_0$ can thus be interpreted as the characteristic Larmor radius of particles with $\gamma\beta\sim \sigma$ in the field equal to the background value; the size of the box is chosen large enough to contain many gyro-orbits of the most energetic particles. Our simulations have a resolution of $d_0=c/\omega_{p0}=5\Delta x$, with $d_0$ and $\omega_{p0}$ being the skin-depth and the plasma frequency of the background plasma; this means that the size of the box is $24000\times 13000$ cells. The particle distribution function is sampled with $5$ particles per cell (PPC) in the upstream, with significantly more in the reconnection layer; the results are virtually unchanged for higher PPC. Deposited currents are smoothed with $8$ passes of a digital filter with a stencil $(1/4,1/2,1/4)$ at each timestep.

To model synchrotron cooling, our simulations also include the radiative drag force in the \cite{1975ctf..book.....L} formulation:
\begin{equation}
\label{eq:sync-force}
\begin{gathered}
  \bm{f}_{\rm rad} =
  \frac{\sigma_T}{4\pi} \left(\bm{\kappa} - \gamma^2\tilde{B}_\perp^2\bm{\beta}\right),~~~\text{where}\\
  \bm{\kappa} = \left(\bm{E} + \bm{\beta}\times\bm{B}\right) \times \bm{B} + \left(\bm{\beta}\cdot\bm{E}\right)\bm{E},\\
  \tilde{B}_\perp^2 = \left(\bm{E} + \bm{\beta}\times\bm{B}\right)^2 - \left(\bm{\beta}\cdot\bm{E}\right)^2,
\end{gathered}
\end{equation}
assuming a particle moves with a four-velocity $\gamma\bm{\beta}$ in the background electromagnetic fields, $\bm{E}$ and $\bm{B}$. To be able to properly model the effect of the drag force at realistic timescales, we upscale its magnitude introducing the dimensionless parameter, $\gamma_{\rm rad}^{\rm syn}$, discussed in sec.~\ref{sec:rad_drag}. The magnitude of the force is scaled in such a way, that $|\bm{f}_{\rm rad}| = 0.1 B_0 |e|$ for $\gamma=\gammarad$, $|\bm{B}|=B_0$, $|\bm{E}|=0$, and $\bm{\beta}\perp \bm{B}$ (with $B_0$ being the strength of the upstream magnetic field). The ratio of $\gammarad/\sigma$ determines the regime of the synchrotron cooling, with $\gammarad/\sigma \lesssim 1$ corresponding to the dynamically strong cooling regime.

\subsection{Results}

To start the reconnection process we remove the pressure balance in the middle of the box, which triggers the formation of two magnetic islands propagating away from the center of the box. After about one light-crossing time, fast plasmoid-mediated magnetic reconnection is triggered across the whole current layer.

\begin{figure*}[p]
    \centering
    \includegraphics[width=\textwidth,trim={0 0 0 0},clip]{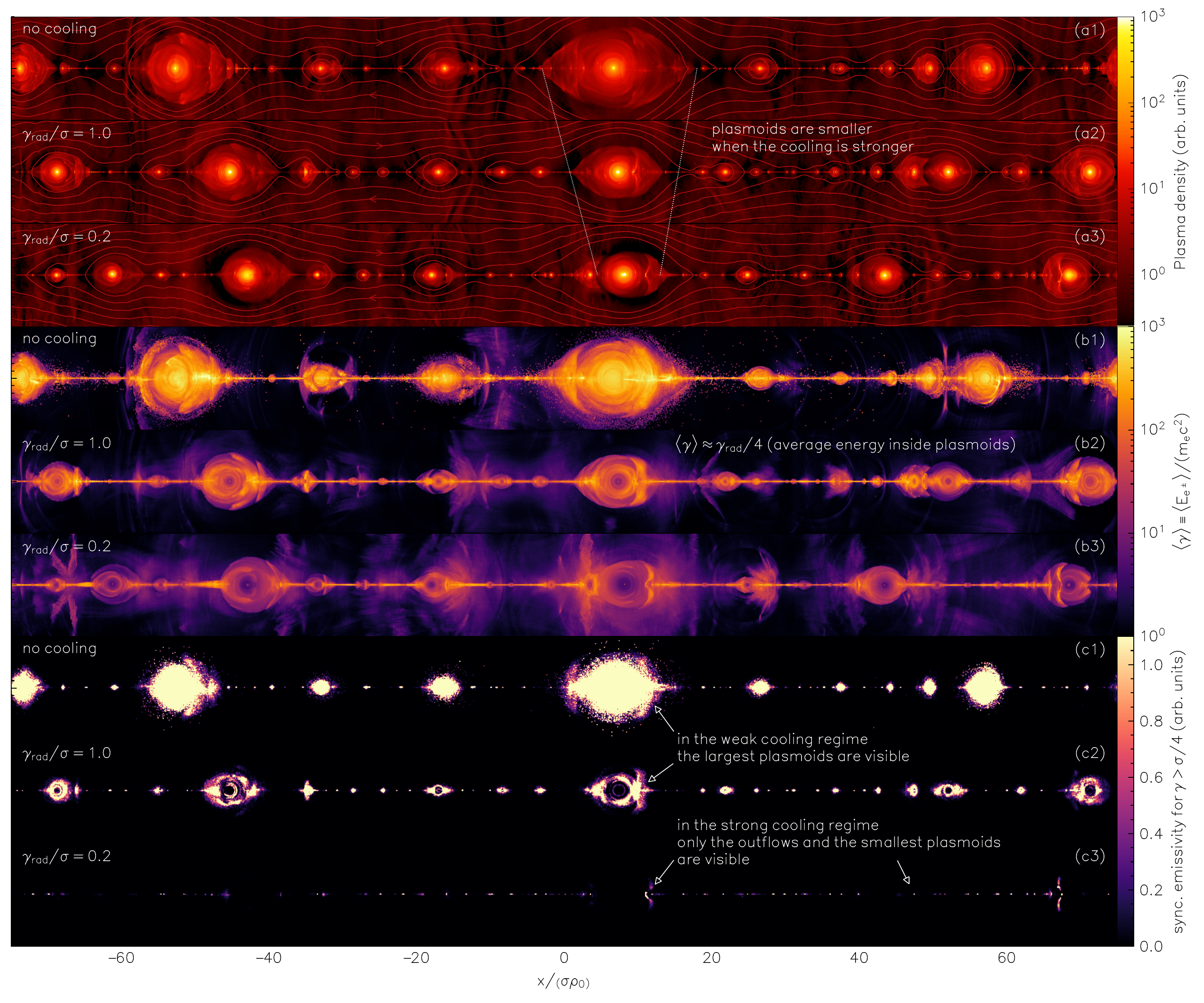}
    \caption{
    Snapshots taken at the same exact timestep from three different 2D simulations of the magnetic reconnection with varying synchrotron cooling strength. Panels from top to bottom show: (a1-a3) the pair-plasma density overplotted with magnetic field lines, (b1-b3) the mean Lorentz factors of particles, (c1-c3) synchrotron emissivity, $\int d\gamma f_{e^\pm}\gamma^2\tilde{B}_\perp^2$, for particles with $\gamma>\sigma/4$. In all three simulations, the upstream plasma magnetization is $\sigma = 10^3$. The ratio $\gammarad/\sigma$ quantifies the strength of synchrotron cooling with smaller values corresponding to stronger cooling. As the cooling strength increases, plasma inside the plasmoids efficiently radiates away the excess momentum perpendicular to the magnetic field. As a result, plasmoids in the strongly cooled case become more compressed (see a1-a3), while the mean energy of the particles inside plasmoids corresponds to the radiation limit, $\langle\gamma\rangle\approx\gammarad/4$. When the cooling is weak, high-energy particles inside plasmoids contribute the most to the synchrotron emission (c1-c2), while in the strong cooling case only the smallest plasmoids and the relativistic outflows in the current sheet are visible. The $x$-coordinate in these snapshots is measured in units of the Larmor radius, $\rho_L$ of particles with energies $\sim \sigma m_e c^2$, i.e., $\rho_L\approx \sigma\rho_0$, where $\rho_0=m_e c^2/(|e|B_0)$, with $B_0$ being the magnetic field strength in the far upstream. 
    }
    \label{fig:2dsnapshot}
\end{figure*}

In Figure~\ref{fig:2dsnapshot} we plot three different quantities for three simulations with varying cooling strength ($\gammarad/\sigma$). Panels a1-a3 show the overall pair-plasma density in units normalized to the upstream density. In these panels, one can clearly see the main features of the reconnecting current sheet: the plasmoids -- circular magnetic structures containing the accelerated plasma, and the x-points in between them, where the magnetic field vanishes and the main energy dissipation takes place. Plasmoids, which travel along the current sheet, adiabatically grow and merge with each other, are held intact via the balance between the magnetic and plasma pressure. From panels a1-a3 one can clearly see that when the synchrotron cooling ``removes'' some of that plasma pressure, densities (and sizes) of the plasmoids have to adjust to maintain the proper balance.

In panels b1-b3 we plot the mean energy of particles in each simulation cell, normalized to $m_e c^2$. From these three panels, it is evident that the dimensionless parameter $\gammarad$ controls the relativistic temperature inside the plasmoids. As the cooling strength increases, i.e., the ratio $\gammarad/\sigma$ drops, the temperature inside the plasmoids decreases, while the density increases, retaining the pressure (the product of the two) roughly constant. We see no evidence of the variation in the plasmoid magnetic field strength for different cooling regimes.

In panels c1-c3 of Figure~\ref{fig:2dsnapshot} we show the total synchrotron emissivity (synchrotron power radiated from a unit volume) for the highest-energy photons. This quantity is defined as $\int d\gamma f_e^\pm \gamma^2\tilde{B}_\perp^2$, where $f_e^\pm$ is the distribution function of high-energy particles (with $\gamma > \sigma/4$), and $\tilde{B}_\perp$ is the effective magnetic field component perpendicular to the direction of the particle velocity, defined in eq. \eqref{eq:sync-force}.
When there is no synchrotron cooling, or when the cooling is dynamically weak (i.e., the regime least applicable to the reconnection layer in the magnetosphere of M87*), plasmoids carry most of the high-energy particles. These particles lose their energies at timescales much longer than the characteristic injection timescale from the current sheet, and the plasmoids get rapidly replenished with fresh accelerated plasma. In this regime most of the high-energy photons are thus produced inside the largest plasmoids (c1-c2). In the strong cooling case, on the other hand, the highest-energy particles are rapidly cooled after they leave the acceleration sites (x-points). Because of that in panel c3 we can only observe the smallest (youngest) plasmoids, as well as the relativistic outflows in the sheet carrying the freshly accelerated pairs, as sources of energetic synchrotron photons. 

In the radiatively efficient regime, $\gammarad/\sigma \lesssim 1$, a fraction which we will denote $\chi^2$, of the dissipated magnetic energy inside the current layer, $p_{\rm diss}\approx |e| E_{\rm rec} c$, is radiated via the synchrotron mechanism: $p_{\rm syn}\approx 2\sigma_T c \langle\gamma^2 \tilde{B}_\perp^2/(8\pi)\rangle=\chi^2 p_{\rm diss}$ (both powers are computed per particle). Here $E_{\rm rec}\approx \beta_{\rm rec} B_0$ is the accelerating electric field in the layer. Using the definition of $\gammarad$ one can find: $\langle\gamma^2 \tilde{B}_\perp^2\rangle\approx \chi^2 \left(\gammarad\right)^2 B_0^2$, a formula that was used in sec.~\ref{sec:syncLuminosity} when computing the total radiated synchrotron power. To find the value of $\chi$ we compare four different simulations with varying cooling strength: one with weak cooling ($\gammarad/\sigma= 2.5$), one with moderate cooling ($\gammarad/\sigma=1$), and the other two with strong cooling ($\gammarad/\sigma= 0.2\text{,~}0.5$). The values for the dimensionless parameter $\chi$ measured from simulations are shown in Table~\ref{tab:results}; the results indicate that for a wide range of values of $\gammarad$ our approximate analytic argument of $\chi$ being constant holds rather well (we inspect this more closely in the appendix \ref{app:sync_lum}).\footnote{Varying $\gamma_{\rm rad}$ by an order of magnitude is equivalent to varying the $B$-field by two orders of magnitude.} We also cite the values of $\kappa\equiv \langle\gamma\rangle/\gammarad$, which was used to estimate the effective temperature of the current sheet. And again, varying the cooling strength within an order of magnitude only marginally affects this parameter, which is close to $0.1\text{-}0.3$.

\begin{table}[h!]
\centering
\setcellgapes{3pt}
\makegapedcells
\begin{tabular}{c!{\vrule width 0.1em} c c c c c}
    \hline
        $\gammarad/\sigma$ &
        $\beta_{\rm rec}$ & 
        $\chi$&
        $\kappa$&
        $p$&
        $\gamma_c$
    \\
    \specialrule{.1em}{.05em}{.05em}
        $\infty$ & $0.25$   & --     & --     & $1.1$        & $2830$
    \\
        $2.5$    & $0.28$   & $0.24$ & $0.10$ & $1.5$        & $1440$
    \\
        $1.0$    & $0.28$   & $0.30$ & $0.17$ & $1.5$        & $1100$
    \\
        $0.5$    & $0.28$   & $0.33$ & $0.22$ & $1.6$, $2.4$ & $1340$
    \\
        $0.2$    & $0.28$   & $0.33$ & $0.29$ & $1.6$, $2.4$ & $1100$
    \\
    \hline
\end{tabular}
\caption{Characteristic values of physical parameters that have observational significance measured directly from our simulations with different cooling strengths. The first column $\gammarad/\sigma$ indicates the cooling strength, with the infinity corresponding to the run without radiation; $\beta_{\rm rec}$ is the time-averaged reconnection rate (normalized to $c$); the ratio $\chi\equiv \langle\gamma^2 \tilde{B}_\perp^2\rangle^{1/2}/\left(\gammarad B_0\right)$ provides an estimate for the total radiated synchrotron power; $\kappa\equiv\langle\gamma\rangle/\gammarad$ correlates the effective temperature with the cooling burnoff limit; $p$ and $\gamma_c$ are the best fit power-law index and the cutoff energy for particle distribution functions $f_e^{\pm}\propto \gamma^{-p}e^{-\gamma/\gamma_c}$. In panels where we cite two values of $p$ we fit a broken power-law with an ankle near $\gamma\approx \gammarad$.}
\label{tab:results}
\end{table}

We also directly measure the reconnection rate by evaluating the component of the $\bm{E}\times\bm{B}/|\bm{B}|^2$ velocity in the upstream, pointing towards the current sheet (in Figure~\ref{fig:2dsnapshot} that corresponds to the $\pm y$ direction). We find that the strength of the synchrotron cooling is unimportant in determining the reconnection rate, with only marginal variations likely caused by the intermittency. The value of the rate itself is close to $\beta_{\rm rec}\sim 0.25\text{-}0.28$, in agreement with what has been found by \cite{2015ApJ...806..167G}, and \cite{sironi2016} in 2D simulations.

\begin{figure}[htb]
    \centering
    \includegraphics[width=1.05\columnwidth]{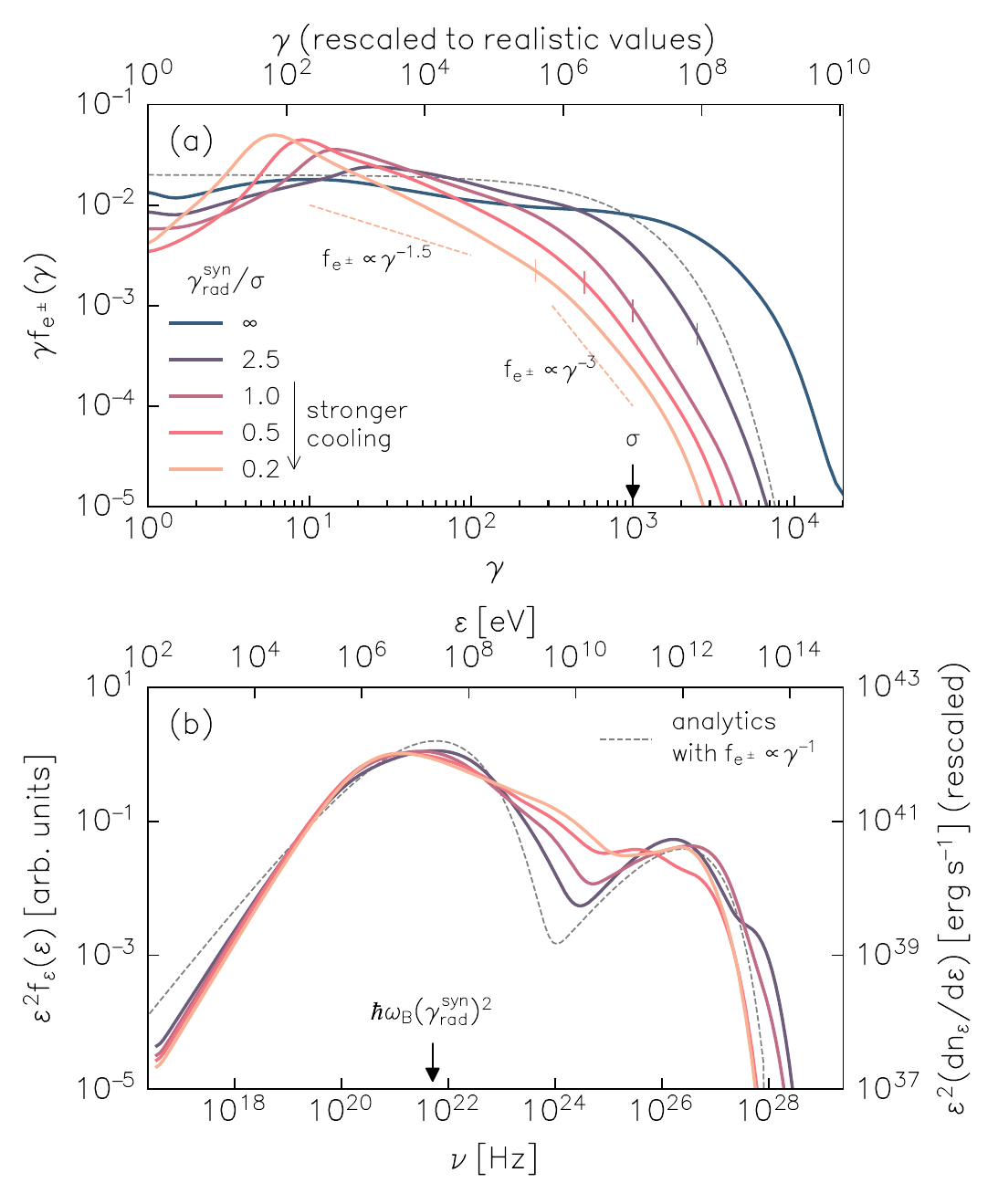}
    \caption{
        Panel (a): The time-averaged distribution function of leptons after one light-crossing time of the simulation. Runs with varying cooling strength, $\gammarad/\sigma$, are shown with different colors, while the magnetization, $\sigma$, is fixed for all the runs. Panel (b): \NEW{time-averaged} synchrotron and IC spectra based on the simulation particles. The synchrotron spectrum (peaked near $\hbar\omega_B\left(\gammarad\right)^2$) is computed on-the-fly, and includes the self-consistent values for $\tilde{B}_\perp$. The IC spectrum (peaked near TeV) is computed in post-processing, assuming an isotropic soft photon background, with the particle Lorentz factors rescaled to realistic values. Dashed lines on \NEW{both panels show the fiducial case of $f_{e^\pm}\propto \gamma^{-1}e^{-\gamma/\gamma_c}$ (the cutoff is fixed at $\gamma_c=\sigma$), which is expected to be applicable for $\sigma \approx 10^7$ in the M87 case, and the analytic prediction for the corresponding synchrotron and IC spectrum.}
    }
    \label{fig:prtlspec}
\end{figure}

Particle-in-cell simulations also enable us to directly measure both the energy distribution of pairs and the spectra of both synchrotron and IC photons from first principles (shown in Figure~\ref{fig:prtlspec}). Particle distribution functions for different values of the cooling strength, $\gammarad/\sigma$, are shown in panel (a) with different colors. Deduced power-law indices and cutoff energies for the best fit power-law (or double-power-law) of the form $f_{e^\pm}\propto\gamma^{-p}e^{-\gamma/\gamma_c}$ (or $\propto\gamma^{-p_1}$ for $\gamma<\gammarad$, and $\propto\gamma^{-p_2}e^{-\gamma/\gamma_c}$ for $\gamma\gtrsim\gammarad$) are shown in the Table~\ref{tab:results}. 

In the weak-to-marginal cooling regime, when $\gammarad/\sigma\gtrsim 1$, the distribution of pairs is best described by a single power-law, with an index varying between $1$ and $1.5$ \NEW{(which justifies the values used in our analytical model in the previous section)}. When the cooling becomes stronger, we see the second power law generated beyond $\gamma\gtrsim\gammarad$ with a slightly steeper index of $\approx 2.5$. In all of the cases, however, except for the uncooled regime, the energy cutoff is strictly controlled by the value of $\sigma$, since the highest-energy pairs are produced in the x-points where there is no synchrotron cooling. In the uncooled case, which is unrealistic for our astrophysical scenario, the spectrum extends much further than $\sigma$ due to various secondary acceleration channels \citep{2018MNRAS.481.5687P, 2021ApJ...912...48H, 2021ApJ...922..261Z}. 

In Figure~\ref{fig:prtlspec}b we show both the synchrotron spectra computed on-the-fly, as well as the IC spectra with an assumed soft isotropic background radiation computed in post-processing. The peaks of the synchrotron emission in all of the cases correspond to $\hbar\omega_B\left(\gammarad\right)^2$, \NEW{which agrees with the earlier predictions by \cite{2014ApJ...780....3U}.} The peaks for the IC signal are self-consistently dictated by the energies of leptons. There are several important differences to the analytic expectation (shown with a dashed line). First, at lower energies our simulation underpredicts the synchrotron flux. This is likely caused by a combination of two reasons: (a) limited scale separation leading to smaller number of low-energy particles in the sheet (we only consider radiation from particles in the current layer), and (b) in 2D, plasmoids, where most of the radiation takes place, are compressed, and magnetic field strengths are stronger than the upstream, which can yield an on average marginally higher radiation frequency. Also notice, that at higher frequencies, $\varepsilon\gtrsim \hbar\omega_B\left(\gammarad\right)^2$, in the synchrotron spectrum we see slightly elevated luminosity, especially for the strongest cooling cases. This is due to the fact that in the strong cooling regime, particles beyond $\gamma\gtrsim\gammarad$ cool very rapidly. 
This makes the spectrum more intermittent, and, when averaged over time, results in more flux at high energies \citep[cf.][]{2019ApJ...877...53H}. To see this, let us estimate the characteristic maximum photon energy, or the cutoff energy. We can use the peak synchrotron energy formula: $E_{\rm cut}\approx (3/2)\hbar \gamma_{\rm max}^2 (|e| \tilde{B}_\perp)/(m_e c^2)$, where $\gamma_{\rm max}\approx \sigma$, and $\tilde{B}_\perp$ is the perpendicular component of the magnetic field for the highest-energy particles. Since $\sigma \gtrsim \gammarad$ (strong cooling regime), particles with Lorentz factors close to $\gamma_{\rm max}$ rapidly lose their energy via synchrotron radiation, emitting the perpendicular component of their momenta. Because of this, the effective $\tilde{B}_\perp$ is smaller than the value of the upstream field, and may be approximated as $\gamma_{\rm max}\tilde{B}_\perp \approx \gammarad B_0$ (see also appendix~\ref{app:sync_lum}). We may then simplify the expression for the spectral cutoff: $E_{\rm cut}\approx (3/2)\hbar \omega_B \left(\gammarad\right)^2 \left(\sigma/\gammarad\right)\approx 16~\text{MeV}\left(\sigma/\gammarad\right)$. We can then conclude that the spectral cutoff for the synchrotron emission is at higher energies for the stronger cooling regimes. 



\vspace{0.5cm}

In our simulations, we had to downscale some of the physical parameters to be able to tackle the problem computationally. In particular, the expected magnetization parameter of the upstream plasma, $\sigma$, during a flare from the magnetosphere of M87* is expected to be close to $10^7$ (see eq. (\ref{eq:sigma_th})), whereas in our simulations we employed $\sigma=10^3$. At the same time the synchrotron limit, $\gammarad$, was downscaled from about $10^6$ to $200$. In light of this, there might be a concern that some of the arguments we have made in this section might not extrapolate when the problem is rescaled to realistic parameters. First of all, it is important to notice that we still retain the strong-cooling hierarchy, $\gammarad<\sigma$, which conserves the balance between the competing mechanisms (acceleration and cooling) when the dimensionless values are downscaled. On top of that, employing $\sigma\sim 10^3$ allows us to have at least a decade of scale separation between $\sigma$, $\gammarad$, and $u_\mathcal{A}/c\sim \sqrt{\sigma}$ (Alfv\'enic four-velocity). We also do not expect the overall dynamics of the layer to change drastically when $u_\mathcal{A}/c\gg 1$, as for these values the Alfv\'enic three-velocity, which dictates the characteristic velocities of the flow in the sheet, approaches the speed of light.

Our simulations are two-dimensional, while the real current sheet in the vicinity of the black hole is inherently 3D. In particular, 3D reconnection has been demonstrated \citep[see][]{2021ApJ...922..261Z} to enable a fast acceleration channel for particles beyond $\gamma\sim\sigma$, which in 2D operates at a much slower rate (\citealt{2018MNRAS.481.5687P,2021ApJ...912...48H}). Ironically, the peculiarity of radiative reconnection helps us in this scenario, since in this regime any secondary acceleration channel outside the x-points is essentially forbidden due to the strong cooling, that vanishes only in these microscopic regions comparable in size to the plasma skin-depth. 

\section{Discussion}
\label{sec:discussion}

The black hole in the center of the M87 galaxy is a perfect target to study not only the dynamics of the large-scale accretion flows but also the extreme plasma-physical processes that intermittently occur in the near vicinity of its event horizon. In particular, long timescales associated with the variability of its core as well as relatively high luminosity makes it a prime target for the observations of TeV flares. At the same time, independent measurements of the magnetic field strength and constraints on the jet power make analytic estimations well-constrained and a lot less ambiguous. While for these reasons the focus of the current paper has been on the M87* black hole, it is important to note that the results can be extrapolated to any system which transiently hosts a magnetospheric, radiatively efficient, reconnecting current sheet (such as potentially Sgr A* or Cen A, and even young energetic pulsars, such as the Crab pulsar; see, e.g., \citealt{1996A&A...311..172L}).

\subsection{Summary}

In this paper, we have shown that large-scale magnetic reconnection, occurring in the episodically forming current sheets in magnetically arrested accretion flows around black holes, can \NEW{power bright multiwavelength flares extending to TeV energies}. We have shown that the reconnecting sheet can dissipate a significant fraction of the jet power on timescales controlled by the universal reconnection rate: $\beta_{\rm rec}\approx 0.1\text{-}0.3$; see eq.~\eqref{eq:sync_luminosity}. While the efficient dissipation of the Poynting flux during magnetic reconnection is not a novel concept,\footnote{For example, the Crab pulsar has been long known to dissipate its spin-down power in the outer-magnetospheric current sheet.} plasma and magnetic field parameters that control the dissipation determine the observational footprint of the process in a wide range of wavelengths. 

Our findings indicate that synchrotron radiation is the main cooling mechanism for the leptons, through which the dissipated magnetic energy is converted into radiation. The synchrotron photons from the accelerated plasma, which peak at around a few tens of MeV, have finite optical depth and are thus susceptible to abundant pair production. This process in turn loads the current sheet with electron-positron pairs; the expected resulting pair-multiplicity with respect to the Goldreich-Julian density is given by eq.~\eqref{eq:multiplicity}, and its estimated value is rather large: $\sim 10^7$. 

We have shown that the expected magnetization parameter in the upstream of the reconnecting current sheet, which determines the available magnetic energy per lepton, is large: $\sigma\approx 10^7$, see eq.~\eqref{eq:sigma_th}. Despite strong synchrotron cooling, reconnection can accelerate pairs to energies close to few-to-tens of TeV,\footnote{\NEW{If there is an additional (unknown) efficient mechanism for plasma loading, the actual magnetization in the current sheet near the M87 black hole may be smaller than our calculations suggest, i.e., $\sigma \lesssim \gammarad$.  In this case, particles can additionally accelerate via 3D secondary mechanisms after they leave the x-points (see, e.g., \citealt{2021ApJ...922..261Z, chernoglazovinprep}). Their energies will eventually be limited by the burnoff limit, $\gamma m_e c^2 \lesssim \gammarad m_e c^2$, i.e., above TeV, which still allows production of TeV emission by the IC scattering.}} or $E_{e^\pm}\approx \sigma m_e c^2$. The highest-energy pairs can then Compton-scatter both the soft photons from the disk and the synchrotron photons from the sheet, producing the very-high-energy TeV component of the emission. Our estimations suggest that the upscattering of the soft thermal photons produced in the bulk of the accretion flow is the most efficient way of producing a TeV signal, with a predicted peak luminosity of $0.01\text{-}0.1\%$ of the jet power, or about $10^{39}\text{-}10^{41}$ erg/s.

We have also verified some of our assumptions using first-principles 2D particle-in-cell (PIC) simulations of reconnecting current sheets with synchrotron cooling included self-consistently. In particular, our simulations support the claim that the high-energy cutoff of the distribution of particles is not sensitive to the synchrotron cooling strength (see table~\ref{tab:results}), since the particle acceleration sites (x-points) have zero magnetic field strength, and are thus not susceptible to synchrotron cooling. On top of that, both the power-law index (which we found to be close to $p\approx 1.5$) and the reconnection rate ($\beta_{\rm rec}\approx 0.28$) are also insensitive to the synchrotron cooling, as long as the cooling is strong. 

\subsection{Observational implications}

Using particle distributions from our simulations we have generated both synchrotron and IC spectra (shown in fig.~\ref{fig:prtlspec}b), with particle Lorentz factors rescaled to realistic values. For the weak cooling regime our generated spectra overall match analytic predictions; both the spectral peak and the cutoff of the synchrotron component are determined by the value of $\gammarad$, and are thus close to $E_{\rm peak}\approx \hbar\omega_B\left(\gammarad\right)^2 \approx 20$ MeV. However, in the strong cooling regime while the peak energy is still set by the value of $\gammarad$, the spectral cutoff is higher $E_{\rm cut}\approx E_{\rm peak}(\sigma/\gammarad)$. This intermittent synchrotron excess at higher energies leads to a much less eminent gap between the synchrotron component (between few-to-ten MeV and GeV) and the inverse-Compton component (above $100$ GeV), with the overall spectrum being more continuous.

GRMHD simulations indicate that during the formation of the current sheet the accretion flow in the magnetospheric region where reconnection takes place is repelled by the magnetic stresses beyond roughly $w\sim 10 r_g$. As a consequence, the radio to near-IR flux from the disk is expected to dim significantly during the flare \citep{jiainprep}. In sec.~\ref{sec:sec_sync} we discussed the possible emergence of a millimeter-to-infrared synchrotron signal from the secondary pairs produced in the reconnection upstream (shown with the pink in Figure~\ref{fig:m87_predict}). The luminosity of this signal is unlikely to exceed $0.1\%\text{-}0.01\%$ of the primary synchrotron counterpart (or about $10^{38}\text{-}10^{39}$ erg/s), as the pair production optical depth, $\tau_{\gamma\gamma}\sim 10^{-4}$, is rather small. Distinguishing it from the dimmed background emission of the accretion flow will thus be challenging, however, a possible feature one should be looking for is \emph{enhanced values of the polarization degree in near-IR during the flare}. The synchrotron radiation from secondary pairs is produced in the reconnection upstream (i.e., at the jet base), where the magnetic field is coherent, as the turbulent accretion flow would have been repelled.\footnote{Another challenge in detecting this component would be the synchrotron self-absorption. The pair-filled environment around the black hole of a size $w\approx 10 r_g$ will be optically thick for energies below $\sim 10^{-4}\text{-}10^{-3}$ eV ($\sim 100\text{-}1000$ GHz). At these frequencies, most of the absorption is caused by the non-thermal secondary pairs with energies $\sim 10\text{-}100~m_e c^2$.} To summarize, we expect that during the TeV flare, the emission of the disk below eV should dim, as the disk is repelled, and the reconnection region is evacuated, while the polarization degree in the same frequency range can increase, owing to the contribution from the synchrotron emission of secondary pairs.


Let us also briefly discuss the temporal characteristics of the flare at different wavelengths. The dimensionless reconnection rate, $\beta_{\rm rec}$, controls both the timescale of the eruption during which the magnetic energy dissipation takes place, as well as its peak synchrotron \NEW{energy}; $\varepsilon_c\approx (3\beta_{\rm rec}/\alpha_F)m_e c^2$, see eq.~\eqref{eq:sync_luminosity}. In the relativistic collisionless regime, studied using PIC simulations for almost a decade, the value of the reconnection rate varies between $0.1\text{-}0.3$ depending on the magnetization parameter, the dimensionality of the simulation, and the boundary conditions (our 2D simulations find values close to $0.25\text{-}0.3$, as shown in the Table~\ref{tab:results}; 3D simulations tend to have smaller values, \citealt{2015ApJ...806..167G}). This prediction is an order of magnitude faster than what has been found by relativistic MHD simulations \citep{ripperda2019}. For the largest current sheets, 3D GRMHD simulations show a flux decay period, i.e., a reconnection event that can power a flare, of $\sim100 r_{\rm g}/c$ \citep[]{Ripperda2022}, which is approximately 30 days for M87*. Due to an order of magnitude faster reconnection rate we expect the flux decay period, and hence the characteristic flare time scale in collisionless plasma to be shorter \citep{Bransgrove2021}, which aligns well with observational flare durations of the order of few days for M87*.\footnote{Note, that the exact duration of the observed flare will also be affected by the effects of beaming and gravitational lensing.}
The magnetic compactness of the inner region near the horizon is significant: $l_B=\sigma_T U_B w/(m_e c^2)\approx \text{few}$ \citep{Beloborodov2017}, with $w\approx 10 r_g$ being the length of the current sheet. This corresponds to the cooling timescale of $t_{\rm cool} \approx (w/c) /(\gamma l_B)$ for particles with a Lorentz factor $\gamma$. Since the cooling time of even the marginally relativistic particles is short, the variability on timescales comparable to $\sim r_g/c$ can only be expected from global effects, such as the beaming due to the orientation of the current sheet. At lower energies, the flare can also have shorter intermittency caused by the dynamics of relativistic outflows, the growth and collisions of plasmoids, and particle acceleration and cooling. For example, in the radio range, which is expected to be produced by the particles with Lorentz factors of the order of $\gamma\sim 100$, the variability timescale can be comparable to $0.1 r_g/c$ (few hours for M87*). At higher energies the same estimate yields $t_{\rm int}\approx t_{\rm cool}(\hbar\omega_B\gamma^2=\varepsilon_s)\approx 10~\text{min}\cdot(\varepsilon_s/\text{eV})^{-1/2}$. While this short-timescale intermittency will be virtually undetectable at the peak energies of MeV (around a second), its detection is possible in the optical to X-ray band, where we expect the characteristic cooling timescale to be around a few tens of seconds to minutes.  

\section*{Acknowledgements}
H.H. would like to express gratitude to Lorenzo Sironi, and Benjamin Crinquand for insightful suggestions and thorough zremarks. B.R. would like to thank Sera Markoff and Joey Neilsen for useful discussions and suggestions. \NEW{The authors would also like to thank the anonymous referee for a detailed report, which helped to significantly improve the quality of the presentation.} The computational resources and services used in this work were partially provided by facilities supported by the Scientific Computing Core at the Flatiron Institute, a division of the Simons Foundation; and by the VSC (Flemish Supercomputer Center), funded by the Research Foundation Flanders (FWO) and the Flemish Government – department EWI. This research is part of the Frontera (\citealt{Frontera}) computing project at the Texas Advanced Computing Center (LRAC-AST21006). Frontera is made possible by National Science Foundation award OAC-1818253. The authors are pleased to acknowledge that the work reported in this paper was performed using the Princeton Research Computing resources at Princeton University which is consortium of groups led by the Princeton Institute for Computational Science and Engineering (PICSciE) and Office of Information Technology's Research Computing.
B.R. is supported by a Joint Princeton/Flatiron Postdoctoral Fellowship. A.P. acknowledges support by the NSF grants No. AST-1910248 and PHY-2010145 and NASA grant 80NSSC22K1054. Research at the Flatiron Institute is supported by the Simons Foundation. This research was facilitated by Multimessenger Plasma Physics Center (MPPC), NSF grant PHY-2206607. H.H. was partially supported by the U.S. Department of Energy under contract number DE-AC02-09CH11466. The United States Government retains a non-exclusive, paid-up, irrevocable, world-wide license to publish or reproduce the published form of this manuscript, or allow others to do so, for United States Government purposes.
\vspace{2cm}

\appendix
\section{Pair production efficiency}
\label{app:pp}
\NEW{
In sec.~\ref{sec:theory} we made an order-of-magnitude estimate for the pair production efficiency from a single photon population by assuming that most of the pairs are produced from the photons close to the peak of the energy distribution, thus greatly simplifying calculations. Here we justify this calculation by considering the process more carefully. We assume that the reconnection produces a distribution of synchrotron photons $N_\gamma \tilde{n}_\gamma(\varepsilon)\propto \varepsilon^{-(p+1)/2}\exp{\left[-(2/3)(\varepsilon/\varepsilon_{\rm max})^{1/3}\right]}$ \citep{2007A&A...465..695Z, 2012ApJ...753..176L}, where $N_\gamma$ is the normalization of the distribution function, i.e., the total number of photons, whereas $\int \tilde{n}_\gamma(\varepsilon)d\varepsilon=1$ (the distribution of pairs producing this synchrotron spectrum is assumed to be $\tilde{n}_\gamma\propto\gamma^{-p}\exp{\left[-\gamma/\gamma_{\rm max}\right]}$, where $\varepsilon_{\rm max}=(3/2)\hbar \omega_B \gamma_{\rm max}^2$). We can then estimate the total pair production yield with the following relation \citep{1987MNRAS.227..403S}:
\begin{equation}
\dot{N}_{e^\pm} \approx \frac{\sigma_T c}{V} N_\gamma^2 
                \int d\varepsilon \tilde{n}_\gamma(\varepsilon)
                a_{\gamma\gamma}(\varepsilon),
\end{equation}
where $V$ is the volume of the pair-production region in the reconnection upstream. Here $a_{\gamma\gamma}(\varepsilon)=\eta_p (m_e c^2/\varepsilon)\tilde{n}_\gamma(m_e c^2/\varepsilon)$ is a dimensionless function describing the angle-averaged cross section for the two-photon pair production, and $\eta_p$ is a dimensionless constant that depends on the slope of the photon distribution. Performing the integration for $\varepsilon > m_e c^2$ yields: $\dot{N}_{e^\pm}\approx (\sigma_T c / V) N_\gamma^2 \eta_p \Lambda(\varepsilon_{\rm max})$, where $\Lambda(\varepsilon_{\rm max})\approx \log{(0.6~\varepsilon_{\rm max}/m_e c^2)}\sim 3$ (for $\varepsilon_{\rm max}\approx 20~\text{MeV}\sim 40~m_e c^2$). For a power-law index $p=1$ we can take, $\eta_p= 7/12$ \citep{1987MNRAS.227..403S} to then conclude: $\dot{N}_{e^\pm}\approx 2(\sigma_T c/V) N_\gamma^2$. We can further take $N_\gamma\approx V U_\gamma / \langle\varepsilon_\gamma \rangle$, with $U_\gamma$ being the average energy density of the synchrotron photons, or expressing the same in terms of the outgoing synchrotron luminosity: $N_\gamma\approx L_\gamma(w/c)/\langle\varepsilon_\gamma\rangle$ ($w$ is the characteristic size of the pair-production region), where we can also take $\langle\varepsilon_\gamma\rangle\approx \varepsilon_{\rm max}$.
}

\section{SSC luminosity}
\label{app:ssc}



To obtain the synchrotron self-Compton power for the parameters applicable for M87* we cannot use the classical Thomson or Klein-Nishina approximations, as the characteristic energies of photons that contribute the most to the outgoing Compton-luminosity in the rest-frame of the highest-energy pairs is close to $m_e c^2$. Instead, we may derive an estimate for the luminosity in this regime using the following formula for the total power \NEW{per lepton} radiated by upscattering photons of energies $\varepsilon_0$ by leptons having Lorentz factors $\gamma$ (see, e.g., \citealt{1970RvMP...42..237B}, \citealt{2012ApJ...753..176L}):

\begin{equation}
\begin{split}
\label{eq:full_ic_power}
    &\frac{dW_{\rm SSC}({\varepsilon}_0,\gamma)}{d{\varepsilon}_0 d\gamma dt} =\\
        &\frac{3}{4}\sigma_T c(m_e c^2)^2 n_0
        \frac{d\tilde{n}_0}{d\varepsilon_0}\frac{d\tilde{n}_{e^\pm}}{d\gamma}
        \frac{\Gamma_\varepsilon^2}{(1+\Gamma_\varepsilon)^2}
        \frac{1}{\varepsilon_0}G_0(\Gamma_\varepsilon),
\end{split}
\end{equation}
Here, the distribution functions of photons, $d\tilde{n}_0/d\varepsilon_0$, and pairs, $d\tilde{n}_{e^\pm}/d\gamma$, are normalized to $1$, and $n_0$ is the number density of incoming synchrotron photons in the steady state. In this equation $\Gamma_\varepsilon\equiv 4\varepsilon_0\gamma/(m_e c^2)$, and $G_0(\Gamma_\varepsilon)$ is a slowly varying function of $\Gamma$ with the asymptotes $G_0({\Gamma_\varepsilon})\approx 1/9$ for $\Gamma_\varepsilon \ll 1$, and $G_0(\Gamma_\varepsilon)\propto (1/2)\ln{\Gamma_\varepsilon}-(11/6)$. Integration of the kernel in eq.~\eqref{eq:full_ic_power} over all values of $\varepsilon_0$ and $\gamma$ yields the total SSC power radiated per unit volume. 

If the synchrotron peak energy, $\varepsilon_s$, is close to $\Gamma_\varepsilon=4\varepsilon_s\gamma_c/(m_e c^2)\approx 1$, we may safely use the $\Gamma_\varepsilon\approx 1$ (and $G_0\approx 1/9$) approximation in eq.~\eqref{eq:full_ic_power}. In this case the total power per lepton is equal to
\begin{equation}
    \label{eq:lowpeak_ssc}
    P_{\rm SSC}^{(1)} \approx \frac{1}{48} \sigma_T c (m_e c^2)^2 n_0 \left\langle\frac{1}{\varepsilon_0}\right\rangle
\end{equation}

On the other hand, for extended power-law distributions of pairs, and the synchrotron peak deep in the Klein-Nishina regime, we may no longer directly substitute $\Gamma_\varepsilon=1$, as the main contribution to the total flux comes from the photons with much lower energies than the peak. To simplify integrating eq.~\eqref{eq:full_ic_power}, it is useful to substitute $\varepsilon_0 = \Gamma_\varepsilon / (4\gamma)$, and assume a power-law distribution for pairs, $d\tilde{n}_{e^\pm}/d\gamma\propto \gamma^{-p}$, and synchrotron photons $d\tilde{n}_0 / d\varepsilon_0 \propto \varepsilon_0^{-(p+1)/2}$. Integration over $\varepsilon_0$ may then be substituted by the integration over $\Gamma_\varepsilon$, and the total SSC power then reads

\begin{equation}
    P_{\rm SSC}^{(2)} \propto \int d\gamma d\Gamma_\varepsilon \frac{d\tilde{n}_{e^\pm}}{d\gamma} \frac{\gamma^{(p+1)/2} \Gamma_\varepsilon^{-(p-1)/2}}{(1+\Gamma_\varepsilon)^2}G_0(\Gamma_\varepsilon).
\end{equation}
Integration over $\Gamma_\varepsilon$ can be well approximated by a delta function near $\Gamma_\varepsilon\approx 1$ \NEW{(for values of $1\le p \lesssim 3$)}, to arrive at the following estimate for the radiated power per lepton:
\begin{equation}
    \label{eq:highpeak_ssc}
    P_{\rm SSC}^{(2)} \approx \frac{3}{4}\sigma_T c \langle\gamma^{-(p+1)/2}\rangle (m_e c^2) n_0.
\end{equation}

\section{Average synchrotron power}
\label{app:sync_lum}

In sect \ref{sec:syncLuminosity} we approximated the distribution-averaged synchrotron power $\langle\gamma^2\tilde{B}_\perp^2\rangle$ as $\chi^2\gamma_{\rm rad}^2 B_0^2$, where $\chi$ is a dimensionless parameter of order unity, and $B_0$ is the strength of the upstream magnetic field. Indeed, as shown in table \ref{tab:results}, the $\chi$ parameter varies only marginally for a range of different cooling strengths ($\gamma_{\rm rad}/\sigma$). 

\begin{figure*}[htb]
    \centering    
    \includegraphics[width=\textwidth]{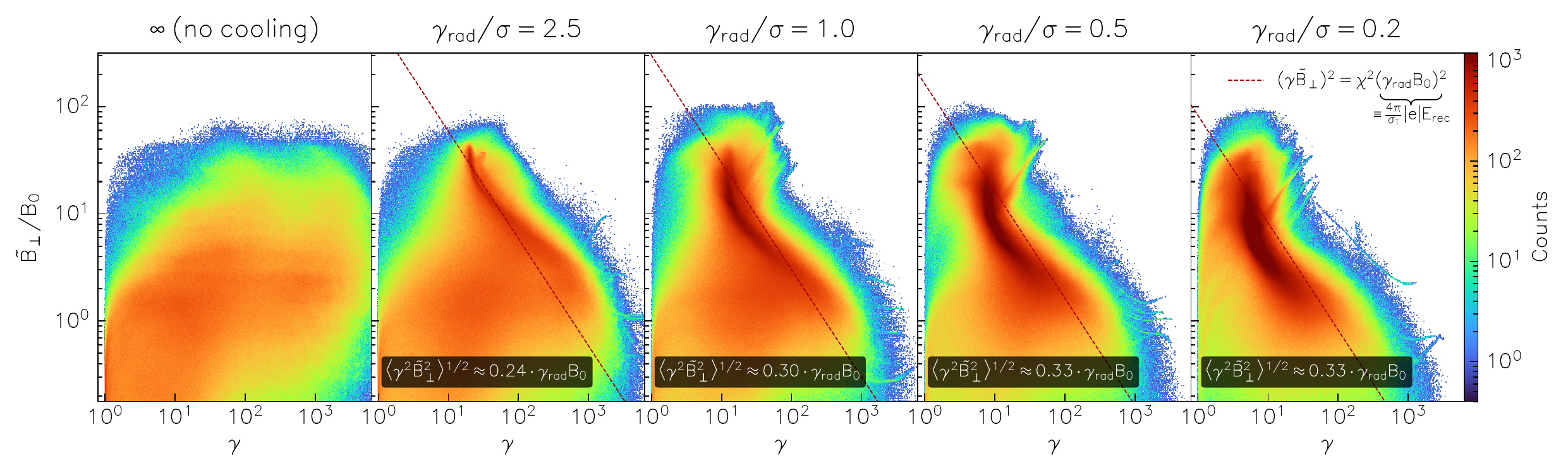}
    \caption{
        Two-dimensional histogram of particles from our 2D reconnection simulations with varying cooling strengths, parametrized by the ratio $\gamma_{\rm rad}/\sigma$ (smaller ratio means stronger cooling). Particles \NEW{are extracted from simulation snapshots at a particular time in the steady-state and} are binned by two parameters: $\tilde{B}_\perp/B_0$ (defined in eq. \eqref{eq:sync-force}), and $\gamma$. At the bottom of each panel where the synchrotron cooling is present we calculate the $\chi$-parameter we used in eq. \eqref{eq:sync_luminosity} to estimate the total synchrotron luminosity. The dashed line roughly shows the strong-to-weak cooling transition (compensated by $\chi^2$). Particles above this line experience synchrotron cooling with strength, $\sigma_T \gamma^2\tilde{B}_\perp^2/(4\pi)$, which is stronger than the characteristic accelerating force from the parallel electric field $|e| E_{\rm rec}$. 
    }
    \label{fig:gammabperp}
\end{figure*}

In figure \ref{fig:gammabperp} we inspect this more closely by plotting 2D histograms for all the particles in our simulations, binning them according to their values of $\gamma$ and $\tilde{B}_\perp$. It is evident, that both the distribution of particles, as well as the compression of plasmoids (i.e., the strength of the magnetic field and, most notably, the density; also evident from figure \ref{fig:2dsnapshot}a) adjust in such a way that the majority of particles are piled near the line corresponding to the cooling-acceleration balance, i.e., $\sigma_T \gamma^2\tilde{B}_\perp^2/(4\pi)\approx |e| E_{\rm rec}$. In the non-cooling case, unsurprisingly, the distribution is a lot more uniform with particles accelerating freely regardless of $\tilde{B}_\perp$. 

In 2D as soon as the energized particles enter the plasmoids, they remain confined inside them until the plasmoids escape the simulation box. In more realistic 3D simulations particles may leave the plasmoids, while traveling in the longitudinal direction (along the plasmoid axis), and the plasmoids themselves can get disrupted due to the kink instability \citep{sironi2014}. Thus, in 3D we do not expect the magnetic field strengths inside the plasmoids to reach values above $B/B_0\sim \text{few}$. The radiation/injection balance discussed above, $\gamma \tilde{B}_\perp\approx \gamma_{\rm rad}^{\rm syn}B_0$, has to still be maintained, and thus the general implications of our 2D simulations will hold even in 3D. However, the details on how particle distribution, as well as the structures of plasmoids, adjust to the cooling-imposed balance condition remains to be seen in future 3D simulations.

\section{Beamed synchrotron self-Compton}
\label{app:beamed-ssc}

In the estimate of the luminosity of the synchrotron self-Compton component, we assumed a random distribution of pitch angles, $\theta$, between the photons and the scattering leptons. Because of this assumption the effective scattering cross section was suppressed by a \NEW{Klein-Nishina} factor $\gamma\varepsilon$ (for photons with energies $\approx \varepsilon$; in this section all the energies are measured in units of $m_e c^2$) with respect to the Thomson cross-section. \NEW{Here, we discuss implications of this assumption. }

\NEW{First, let us briefly reformulate our previous result on the energy yield of the SSC emission in terms of the effective scattering cross-section.} The number of scatterings is proportional to the effective IC cross section, $\sigma_{\rm eff}$, while the energy yield scales with the maximum energy of scattered photons: $P_{\rm SSC}\propto \sigma_{\rm eff}\varepsilon_{\rm max}\approx \sigma_{\rm eff}\gamma_c$. Comparing this with the synchrotron luminosity, $P_{\rm syn}\propto \sigma_T\varepsilon_c$, yields: $P_{\rm SSC}/P_{\rm syn}\approx \sigma_{\rm eff}\gamma/(\sigma_T \varepsilon_c)\ll 1$, where the exact value of $\sigma_{\rm eff}$ depends on the energy distribution of incoming photons. From eq. \eqref{eq:highpeak_ssc} we can see that for the applicable regime it is close to $\sigma_{\rm eff}\approx \sigma_T \gamma_c^{-(p+3)/2}$ (this is due to the fact that most of the outgoing energy is accumulated by upscattering the low-energy synchrotron photons far below the peak, $\varepsilon_c$). This finally yields: $P_{\rm SSC}/P_{\rm syn} \approx 1/(\gamma_c^{(p+1)/2} \varepsilon_c)\ll 1$.

In the more general case the effective cross-section will be determined by the distribution of pitch angles of the incoming synchrotron photons $dn_{\varepsilon e^\pm}/d\theta$: $\sigma_{\rm eff}=\sigma'\varepsilon'/(\gamma\varepsilon) \approx \int_0^\pi [\sigma_\theta(1-\beta\cos{\theta})(dn_{\varepsilon e^\pm}/d\theta)]d\theta$ (primed quantities are measured in the lepton reference frame: $\varepsilon'=\gamma\varepsilon(1-\beta\cos{\theta})$; see, e.g., \citealt{1975ctf..book.....L}). Here $\sigma_\theta\approx \sigma_T$ when $\theta\ll\theta_T\equiv 1/\sqrt{\gamma\varepsilon}$ (Thomson regime), and $\approx \sigma_T/(\gamma\varepsilon)$ -- otherwise (Klein-Nishina regime). If the pitch-angle distribution is uniform, $dn_{\varepsilon e^\pm}/d\theta=1/\pi$, the effective cross section is suppressed by a factor of $1/(\gamma \varepsilon_c)$ (in agreement with our original conclusion for $\gamma\approx \gamma_c$ and $p=1$).

One might expect that since the emission of the synchrotron photons is beamed in the direction of particles, Compton scattering of the same population will be more efficient, as the photons will have smaller energy in the lepton rest frame. In the most optimistic scenario when all the photons are beamed in the direction of particles, i.e., $dn_{\varepsilon e^\pm}/d\theta$ is a delta function in $\theta$, we find $\sigma_{\rm eff}= \sigma_T(1-\beta)\approx \sigma_T/(2\gamma^2)$. In terms of the outgoing luminosity this results in $P_{\rm SSC}/P_{\rm syn}\approx 1/(\gamma_c\varepsilon_c)$. Since the optical depth for the synchrotron photons to Compton scattering in the large-scale reconnecting sheet of the black-hole magnetosphere is finite, the scatterings are not entirely local, and only a fraction of the scatterings can be considered beamed. The actual effective cross section is then $\sigma_T/(2\gamma^2)\lesssim\sigma_{\rm eff}\lesssim \sigma_T/(\gamma\varepsilon)$, which is still $\ll \sigma_T$, and thus $P_{\rm SSC}\ll P_{\rm syn}$.

\bibliography{refs_br1,refs_br3,refs_hh}{}

\begin{thebibliography}{}
\expandafter\ifx\csname natexlab\endcsname\relax\def\natexlab#1{#1}\fi
\providecommand{\url}[1]{\href{#1}{#1}}
\providecommand{\dodoi}[1]{doi:~\href{http://doi.org/#1}{\nolinkurl{#1}}}
\providecommand{\doeprint}[1]{\href{http://ascl.net/#1}{\nolinkurl{http://ascl.net/#1}}}
\providecommand{\doarXiv}[1]{\href{https://arxiv.org/abs/#1}{\nolinkurl{https://arxiv.org/abs/#1}}}

\bibitem[{{Abramowski} {et~al.}(2012){Abramowski}, {Acero}, {Aharonian},
  {Akhperjanian}, {Anton}, {Balzer}, {Barnacka}, {Barres de Almeida},
  {Becherini}, {Becker}, {Behera}, {Bernl{\"o}hr}, {Birsin}, {Biteau},
  {Bochow}, {Boisson}, {Bolmont}, {Bordas}, {Brucker}, {Brun}, {Brun}, {Bulik},
  {B{\"u}sching}, {Carrigan}, {Casanova}, {Cerruti}, {Chadwick}, {Charbonnier},
  {Chaves}, {Cheesebrough}, {Clapson}, {Coignet}, {Cologna}, {Conrad},
  {Dalton}, {Daniel}, {Davids}, {Degrange}, {Deil}, {Dickinson},
  {Djannati-Ata{\"\i}}, {Domainko}, {Drury}, {Dubus}, {Dutson}, {Dyks},
  {Dyrda}, {Egberts}, {Eger}, {Espigat}, {Fallon}, {Farnier}, {Fegan},
  {Feinstein}, {Fernandes}, {Fiasson}, {Fontaine}, {F{\"o}rster},
  {F{\"u}{\ss}ling}, {Gallant}, {Gast}, {G{\'e}rard}, {Gerbig}, {Giebels},
  {Glicenstein}, {Gl{\"u}ck}, {Goret}, {G{\"o}ring}, {H{\"a}ffner}, {Hague},
  {Hampf}, {Hauser}, {Heinz}, {Heinzelmann}, {Henri}, {Hermann}, {Hinton},
  {Hoffmann}, {Hofmann}, {Hofverberg}, {Holler}, {Horns}, {Jacholkowska}, {de
  Jager}, {Jahn}, {Jamrozy}, {Jung}, {Kastendieck}, {Katarzy{\'n}ski}, {Katz},
  {Kaufmann}, {Keogh}, {Khangulyan}, {Kh{\'e}lifi}, {Klochkov}, {Klu{\'z}niak},
  {Kneiske}, {Komin}, {Kosack}, {Kossakowski}, {Laffon}, {Lamanna}, {Lennarz},
  {Lohse}, {Lopatin}, {Lu}, {Marandon}, {Marcowith}, {Masbou}, {Maurin},
  {Maxted}, {Mayer}, {McComb}, {Medina}, {M{\'e}hault}, {Moderski}, {Moulin},
  {Naumann}, {Naumann-Godo}, {de Naurois}, {Nedbal}, {Nekrassov}, {Nguyen},
  {Nicholas}, {Niemiec}, {Nolan}, {Ohm}, {de O{\~n}a Wilhelmi}, {Opitz},
  {Ostrowski}, {Oya}, {Panter}, {Paz Arribas}, {Pedaletti}, {Pelletier},
  {Petrucci}, {Pita}, {P{\"u}hlhofer}, {Punch}, {Quirrenbach}, {Raue},
  {Rayner}, {Reimer}, {Reimer}, {Renaud}, {de los Reyes}, {Rieger}, {Ripken},
  {Rob}, {Rosier-Lees}, {Rowell}, {Rudak}, {Rulten}, {Ruppel}, {Sahakian},
  {Sanchez}, {Santangelo}, {Schlickeiser}, {Sch{\"o}ck}, {Schulz}, {Schwanke},
  {Schwarzburg}, {Schwemmer}, {Sheidaei}, {Skilton}, {Sol}, {Spengler},
  {Stawarz}, {Steenkamp}, {Stegmann}, {Stinzing}, {Stycz}, {Sushch}, {Szostek},
  {Tavernet}, {Terrier}, {Tluczykont}, {Valerius}, {van Eldik}, {Vasileiadis},
  {Venter}, {Vialle}, {Viana}, {Vincent}, {V{\"o}lk}, {Volpe}, {Vorobiov},
  {Vorster}, {Wagner}, {Ward}, {White}, {Wierzcholska}, {Zacharias}, {Zajczyk},
  {Zdziarski}, {Zech}, {Zechlin}, {H.~E.~S.~S. Collaboration}, {Aleksi{\'c}},
  {Antonelli}, {Antoranz}, {Backes}, {Barrio}, {Bastieri}, {Becerra
  Gonz{\'a}lez}, {Bednarek}, {Berdyugin}, {Berger}, {Bernardini}, {Biland},
  {Blanch}, {Bock}, {Boller}, {Bonnoli}, {Borla Tridon}, {Braun}, {Bretz},
  {Ca{\~n}ellas}, {Carmona}, {Carosi}, {Colin}, {Colombo}, {Contreras},
  {Cortina}, {Cossio}, {Covino}, {Dazzi}, {De Angelis}, {De Cea del Pozo}, {De
  Lotto}, {Delgado Mendez}, {Diago Ortega}, {Doert}, {Dom{\'\i}nguez}, {Dominis
  Prester}, {Dorner}, {Doro}, {Elsaesser}, {Ferenc}, {Fonseca}, {Font},
  {Fruck}, {Garc{\'\i}a L{\'o}pez}, {Garczarczyk}, {Garrido}, {Giavitto},
  {Godinovi{\'c}}, {Hadasch}, {H{\"a}fner}, {Herrero}, {Hildebrand},
  {H{\"o}hne-M{\"o}nch}, {Hose}, {Hrupec}, {Huber}, {Jogler}, {Klepser},
  {Kr{\"a}henb{\"u}hl}, {Krause}, {La Barbera}, {Lelas}, {Leonardo},
  {Lindfors}, {Lombardi}, {L{\'o}pez}, {Lorenz}, {Makariev}, {Maneva},
  {Mankuzhiyil}, {Mannheim}, {Maraschi}, {Mariotti}, {Mart{\'\i}nez}, {Mazin},
  {Meucci}, {Miranda}, {Mirzoyan}, {Miyamoto}, {Mold{\'o}n}, {Moralejo},
  {Munar}, {Nieto}, {Nilsson}, {Orito}, {Oya}, {Paneque}, {Paoletti}, {Pardo},
  {Paredes}, {Partini}, {Pasanen}, {Pauss}, {Perez-Torres}, {Persic},
  {Peruzzo}, {Pilia}, {Pochon}, {Prada}, {Prada Moroni}, {Prandini}, {Puljak},
  {Reichardt}, {Reinthal}, {Rhode}, {Rib{\'o}}, {Rico}, {R{\"u}gamer},
  {Saggion}, {Saito}, {Saito}, {Salvati}, {Satalecka}, {Scalzotto}, {Scapin},
  {Schultz}, {Schweizer}, {Shayduk}, {Shore}, {Sillanp{\"a}{\"a}}, {Sitarek},
  {Sobczynska}, {Spanier}, {Spiro}, {Stamerra}, {Steinke}, {Storz}, {Strah},
  {Suri{\'c}}, {Takalo}, {Takami}, {Tavecchio}, {Temnikov}, {Terzi{\'c}},
  {Tescaro}, {Teshima}, {Thom}, {Tibolla}, {Torres}, {Treves}, {Vankov},
  {Vogler}, {Wagner}, {Weitzel}, {Zabalza}, {Zandanel}, {Zanin}, {MAGIC
  Collaboration}, {Arlen}, {Aune}, {Beilicke}, {Benbow}, {Bouvier}, {Bradbury},
  {Buckley}, {Bugaev}, {Byrum}, {Cannon}, {Cesarini}, {Ciupik}, {Connolly},
  {Cui}, {Dickherber}, {Duke}, {Errando}, {Falcone}, {Finley}, {Finnegan},
  {Fortson}, {Furniss}, {Galante}, {Gall}, {Godambe}, {Griffin}, {Grube},
  {Gyuk}, {Hanna}, {Holder}, {Huan}, {Hui}, {Kaaret}, {Karlsson}, {Kertzman},
  {Khassen}, {Kieda}, {Krawczynski}, {Krennrich}, {Lang}, {LeBohec}, {Maier},
  {McArthur}, {McCann}, {Moriarty}, {Mukherjee}, {Nu{\~n}ez}, {Ong}, {Orr},
  {Otte}, {Park}, {Perkins}, {Pichel}, {Pohl}, {Prokoph}, {Ragan}, {Reyes},
  {Reynolds}, {Roache}, {Rose}, {Ruppel}, {Schroedter}, {Sembroski},
  {{\c{S}}ent{\"u}rk}, {Telezhinsky}, {Te{\v{s}}i{\'c}}, {Theiling},
  {Thibadeau}, {Varlotta}, {Vassiliev}, {Vivier}, {Wakely}, {Weekes},
  {Williams}, {Zitzer}, {VERITAS Collaboration}, {Barres de Almeida}, {Cara},
  {Casadio}, {Cheung}, {McConville}, {Davies}, {Doi}, {Giovannini},
  {Giroletti}, {Hada}, {Hardee}, {Harris}, {Junor}, {Kino}, {Lee}, {Ly},
  {Madrid}, {Massaro}, {Mundell}, {Nagai}, {Perlman}, {Steele}, {Walker}, \&
  {Wood}}]{Hess2012}
{Abramowski}, A., {Acero}, F., {Aharonian}, F., {et~al.} 2012, \apj, 746, 151

\bibitem[{{Acciari} {et~al.}(2010){Acciari}, {Aliu}, {Arlen}, {Aune},
  {Beilicke}, {Benbow}, {Boltuch}, {Bradbury}, {Buckley}, {Bugaev}, {Byrum},
  {Cannon}, {Cesarini}, {Chow}, {Ciupik}, {Cogan}, {Cui}, {Dickherber}, {Duke},
  {Finley}, {Finnegan}, {Fortin}, {Fortson}, {Furniss}, {Galante}, {Gall},
  {Gillanders}, {Godambe}, {Grube}, {Guenette}, {Gyuk}, {Hanna}, {Holder},
  {Hui}, {Humensky}, {Imran}, {Kaaret}, {Karlsson}, {Kertzman}, {Kieda},
  {Konopelko}, {Krawczynski}, {Krennrich}, {Lang}, {LeBohec}, {Maier},
  {McArthur}, {McCann}, {McCutcheon}, {Millis}, {Moriarty}, {Ong}, {Otte},
  {Pandel}, {Perkins}, {Pichel}, {Pohl}, {Quinn}, {Ragan}, {Reyes}, {Reynolds},
  {Roache}, {Rose}, {Rovero}, {Schroedter}, {Sembroski}, {Senturk}, {Smith},
  {Steele}, {Swordy}, {Theiling}, {Thibadeau}, {Varlotta}, {Vincent}, {Wagner},
  {Wakely}, {Ward}, {Weekes}, {Weinstein}, {Weisgarber}, {Williams}, {Wissel},
  {Wood}, {Zitzer}, {Harris}, \& {Massaro}}]{Veritas2010}
{Acciari}, V.~A., {Aliu}, E., {Arlen}, T., {et~al.} 2010, \apj, 716, 819

\bibitem[{{Aharonian} {et~al.}(2006){Aharonian}, {Akhperjanian}, {Bazer-Bachi},
  {Beilicke}, {Benbow}, {Berge}, {Bernl{\"o}hr}, {Boisson}, {Bolz}, {Borrel},
  {Braun}, {Brown}, {B{\"u}hler}, {B{\"u}sching}, {Carrigan}, {Chadwick},
  {Chounet}, {Coignet}, {Cornils}, {Costamante}, {Degrange}, {Dickinson},
  {Djannati-Ata{\"\i}}, {Drury}, {Dubus}, {Egberts}, {Emmanoulopoulos},
  {Espigat}, {Feinstein}, {Ferrero}, {Fiasson}, {Fontaine}, {Funk}, {Funk},
  {F{\"u}{\ss}ling}, {Gallant}, {Giebels}, {Glicenstein}, {Goret},
  {Hadjichristidis}, {Hauser}, {Hauser}, {Heinzelmann}, {Henri}, {Hermann},
  {Hinton}, {Hoffmann}, {Hofmann}, {Holleran}, {Hoppe}, {Horns},
  {Jacholkowska}, {de Jager}, {Kendziorra}, {Kerschhaggl}, {Kh{\'e}lifi},
  {Komin}, {Konopelko}, {Kosack}, {Lamanna}, {Latham}, {Le Gallou},
  {Lemi{\`e}re}, {Lemoine-Goumard}, {Lenain}, {Lohse}, {Martin},
  {Martineau-Huynh}, {Marcowith}, {Masterson}, {Maurin}, {McComb}, {Moulin},
  {de Naurois}, {Nedbal}, {Nolan}, {Noutsos}, {Orford}, {Osborne}, {Ouchrif},
  {Panter}, {Pelletier}, {Pita}, {P{\"u}hlhofer}, {Punch}, {Ranchon},
  {Raubenheimer}, {Raue}, {Rayner}, {Reimer}, {Ripken}, {Rob}, {Rolland},
  {Rosier-Lees}, {Rowell}, {Sahakian}, {Santangelo}, {Saug{\'e}}, {Schlenker},
  {Schlickeiser}, {Schr{\"o}der}, {Schwanke}, {Schwarzburg}, {Schwemmer},
  {Shalchi}, {Sol}, {Spangler}, {Spanier}, {Steenkamp}, {Stegmann}, {Superina},
  {Tam}, {Tavernet}, {Terrier}, {Tluczykont}, {van Eldik}, {Vasileiadis},
  {Venter}, {Vialle}, {Vincent}, {V{\"o}lk}, {Wagner}, \& {Ward}}]{Hess2006}
{Aharonian}, F., {Akhperjanian}, A.~G., {Bazer-Bachi}, A.~R., {et~al.} 2006,
  Science, 314, 1424

\bibitem[{{Aharonian} {et~al.}(2007){Aharonian}, {Akhperjanian}, {Bazer-Bachi},
  {Behera}, {Beilicke}, {Benbow}, {Berge}, {Bernl{\"o}hr}, {Boisson}, {Bolz},
  {Borrel}, {Boutelier}, {Braun}, {Brion}, {Brown}, {B{\"u}hler},
  {B{\"u}sching}, {Bulik}, {Carrigan}, {Chadwick}, {Clapson}, {Chounet},
  {Coignet}, {Cornils}, {Costamante}, {Degrange}, {Dickinson},
  {Djannati-Ata{\"\i}}, {Domainko}, {Drury}, {Dubus}, {Dyks}, {Egberts},
  {Emmanoulopoulos}, {Espigat}, {Farnier}, {Feinstein}, {Fiasson},
  {F{\"o}rster}, {Fontaine}, {Funk}, {Funk}, {F{\"u}{\ss}ling}, {Gallant},
  {Giebels}, {Glicenstein}, {Gl{\"u}ck}, {Goret}, {Hadjichristidis}, {Hauser},
  {Hauser}, {Heinzelmann}, {Henri}, {Hermann}, {Hinton}, {Hoffmann}, {Hofmann},
  {Holleran}, {Hoppe}, {Horns}, {Jacholkowska}, {de Jager}, {Kendziorra},
  {Kerschhaggl}, {Kh{\'e}lifi}, {Komin}, {Kosack}, {Lamanna}, {Latham}, {Le
  Gallou}, {Lemi{\`e}re}, {Lemoine-Goumard}, {Lenain}, {Lohse}, {Martin},
  {Martineau-Huynh}, {Marcowith}, {Masterson}, {Maurin}, {McComb}, {Moderski},
  {Moulin}, {de Naurois}, {Nedbal}, {Nolan}, {Olive}, {Orford}, {Osborne},
  {Ostrowski}, {Panter}, {Pedaletti}, {Pelletier}, {Petrucci}, {Pita},
  {P{\"u}hlhofer}, {Punch}, {Ranchon}, {Raubenheimer}, {Raue}, {Rayner},
  {Renaud}, {Ripken}, {Rob}, {Rolland}, {Rosier-Lees}, {Rowell}, {Rudak},
  {Ruppel}, {Sahakian}, {Santangelo}, {Saug{\'e}}, {Schlenker}, {Schlickeiser},
  {Schr{\"o}der}, {Schwanke}, {Schwarzburg}, {Schwemmer}, {Shalchi}, {Sol},
  {Spangler}, {Stawarz}, {Steenkamp}, {Stegmann}, {Superina}, {Tam},
  {Tavernet}, {Terrier}, {van Eldik}, {Vasileiadis}, {Venter}, {Vialle},
  {Vincent}, {Vivier}, {V{\"o}lk}, {Volpe}, {Wagner}, {Ward}, \&
  {Zdziarski}}]{Aharonian2007}
---. 2007, \apjl, 664, L71

\bibitem[{{Aharonian} {et~al.}(2009){Aharonian}, {Akhperjanian}, {Anton}, {de
  Almeida}, {Bazer-Bachi}, {Becherini}, {Behera}, {Benbow}, {Bernl{\"o}hr},
  {Boisson}, {Bochow}, {Borrel}, {Brion}, {Brucker}, {Brun}, {B{\"u}hler},
  {Bulik}, {B{\"u}sching}, {Boutelier}, {Chadwick}, {Charbonnier}, {Chaves},
  {Cheesebrough}, {Chounet}, {Clapson}, {Coignet}, {Dalton}, {Daniel},
  {Davids}, {Degrange}, {Deil}, {Dickinson}, {Djannati-Ata{\"\i}}, {Domainko},
  {Drury}, {Dubois}, {Dubus}, {Dyks}, {Dyrda}, {Egberts}, {Emmanoulopoulos},
  {Espigat}, {Farnier}, {Feinstein}, {Fiasson}, {F{\"o}rster}, {Fontaine},
  {F{\"u}{\ss}ling}, {Gabici}, {Gallant}, {G{\'e}rard}, {Giebels},
  {Glicenstein}, {Gl{\"u}ck}, {Goret}, {G{\"o}hring}, {Hauser}, {Hauser},
  {Heinz}, {Heinzelmann}, {Henri}, {Hermann}, {Hinton}, {Hoffmann}, {Hofmann},
  {Holleran}, {Hoppe}, {Horns}, {Jacholkowska}, {de Jager}, {Jahn}, {Jung},
  {Katarzy{\'n}ski}, {Katz}, {Kaufmann}, {Kendziorra}, {Kerschhaggl},
  {Khangulyan}, {Kh{\'e}lifi}, {Keogh}, {Klu{\'z}niak}, {Kneiske}, {Komin},
  {Kosack}, {Lamanna}, {Latham}, {Lenain}, {Lohse}, {Marandon}, {Martin},
  {Martineau-Huynh}, {Marcowith}, {Maurin}, {McComb}, {Medina}, {Moderski},
  {Moulin}, {Naumann-Godo}, {de Naurois}, {Nedbal}, {Nekrassov}, {Niemiec},
  {Nolan}, {Ohm}, {Olive}, {de O{\~n}a Wilhelmi}, {Orford}, {Ostrowski},
  {Panter}, {Arribas}, {Pedaletti}, {Pelletier}, {Petrucci}, {Pita},
  {P{\"u}hlhofer}, {Punch}, {Quirrenbach}, {Raubenheimer}, {Raue}, {Rayner},
  {Renaud}, {Rieger}, {Ripken}, {Rob}, {Rosier-Lees}, {Rowell}, {Rudak},
  {Rulten}, {Ruppel}, {Sahakian}, {Santangelo}, {Schlickeiser}, {Sch{\"o}ck},
  {Schr{\"o}der}, {Schwanke}, {Schwarzburg}, {Schwemmer}, {Shalchi}, {Sikora},
  {Skilton}, {Sol}, {Spangler}, {Stawarz}, {Steenkamp}, {Stegmann}, {Superina},
  {Szostek}, {Tam}, {Tavernet}, {Terrier}, {Tibolla}, {Tluczykont}, {van
  Eldik}, {Vasileiadis}, {Venter}, {Venter}, {Vialle}, {Vincent}, {Vink},
  {Vivier}, {V{\"o}lk}, {Volpe}, {Wagner}, {Ward}, {Zdziarski}, \&
  {Zech}}]{Aharonian2009}
{Aharonian}, F., {Akhperjanian}, A.~G., {Anton}, G., {et~al.} 2009, \apjl, 695,
  L40

\bibitem[{{Akhiezer} \& {Berestetskij}(1985)}]{1985quel.book.....A}
{Akhiezer}, A., \& {Berestetskij}, V.~B. 1985, {Quantum Electrodynamics}

\bibitem[{{Albert} {et~al.}(2007){Albert}, {Aliu}, {Anderhub}, {Antoranz},
  {Armada}, {Baixeras}, {Barrio}, {Bartko}, {Bastieri}, {Becker}, {Bednarek},
  {Berger}, {Bigongiari}, {Biland}, {Bock}, {Bordas}, {Bosch-Ramon}, {Bretz},
  {Britvitch}, {Camara}, {Carmona}, {Chilingarian}, {Coarasa}, {Commichau},
  {Contreras}, {Cortina}, {Costado}, {Curtef}, {Danielyan}, {Dazzi}, {De
  Angelis}, {Delgado}, {de los Reyes}, {De Lotto}, {Domingo-Santamar{\'\i}a},
  {Dorner}, {Doro}, {Errando}, {Fagiolini}, {Ferenc}, {Fern{\'a}ndez}, {Firpo},
  {Flix}, {Fonseca}, {Font}, {Fuchs}, {Galante}, {Garc{\'\i}a-L{\'o}pez},
  {Garczarczyk}, {Gaug}, {Giller}, {Goebel}, {Hakobyan}, {Hayashida},
  {Hengstebeck}, {Herrero}, {H{\"o}hne}, {Hose}, {Hrupec}, {Hsu}, {Jacon},
  {Jogler}, {Kosyra}, {Kranich}, {Kritzer}, {Laille}, {Lindfors}, {Lombardi},
  {Longo}, {L{\'o}pez}, {L{\'o}pez}, {Lorenz}, {Majumdar}, {Maneva},
  {Mannheim}, {Mansutti}, {Mariotti}, {Mart{\'\i}nez}, {Mazin}, {Merck},
  {Meucci}, {Meyer}, {Miranda}, {Mirzoyan}, {Mizobuchi}, {Moralejo}, {Nieto},
  {Nilsson}, {Ninkovic}, {O{\~n}a-Wilhelmi}, {Otte}, {Oya}, {Paneque},
  {Panniello}, {Paoletti}, {Paredes}, {Pasanen}, {Pascoli}, {Pauss}, {Pegna},
  {Persic}, {Peruzzo}, {Piccioli}, {Prandini}, {Puchades}, {Raymers}, {Rhode},
  {Rib{\'o}}, {Rico}, {Rissi}, {Robert}, {R{\"u}gamer}, {Saggion}, {Saito},
  {S{\'a}nchez}, {Sartori}, {Scalzotto}, {Scapin}, {Schmitt}, {Schweizer},
  {Shayduk}, {Shinozaki}, {Shore}, {Sidro}, {Sillanp{\"a}{\"a}}, {Sobczynska},
  {Stamerra}, {Stark}, {Takalo}, {Tavecchio}, {Temnikov}, {Tescaro}, {Teshima},
  {Torres}, {Turini}, {Vankov}, {Vitale}, {Wagner}, {Wibig}, {Wittek},
  {Zandanel}, {Zanin}, \& {Zapatero}}]{Albert2007}
{Albert}, J., {Aliu}, E., {Anderhub}, H., {et~al.} 2007, \apj, 669, 862

\bibitem[{{Aleksi{\'c}} {et~al.}(2014){Aleksi{\'c}}, {Ansoldi}, {Antonelli},
  {Antoranz}, {Babic}, {Bangale}, {Barrio}, {Gonz{\'a}lez}, {Bednarek},
  {Bernardini}, {Biasuzzi}, {Biland}, {Blanch}, {Bonnefoy}, {Bonnoli},
  {Borracci}, {Bretz}, {Carmona}, {Carosi}, {Colin}, {Colombo}, {Contreras},
  {Cortina}, {Covino}, {Da Vela}, {Dazzi}, {De Angelis}, {De Caneva}, {De
  Lotto}, {Wilhelmi}, {Mendez}, {Prester}, {Dorner}, {Doro}, {Einecke},
  {Eisenacher}, {Elsaesser}, {Fonseca}, {Font}, {Frantzen}, {Fruck}, {Galindo},
  {L{\'o}pez}, {Garczarczyk}, {Terrats}, {Gaug}, {Godinovi{\'c}}, {Mu{\~n}oz},
  {Gozzini}, {Hadasch}, {Hanabata}, {Hayashida}, {Herrera}, {Hildebrand},
  {Hose}, {Hrupec}, {Idec}, {Kadenius}, {Kellermann}, {Kodani}, {Konno},
  {Krause}, {Kubo}, {Kushida}, {La Barbera}, {Lelas}, {Lewandowska},
  {Lindfors}, {Lombardi}, {Longo}, {L{\'o}pez}, {L{\'o}pez-Coto},
  {L{\'o}pez-Oramas}, {Lorenz}, {Lozano}, {Makariev}, {Mallot}, {Maneva},
  {Mankuzhiyil}, {Mannheim}, {Maraschi}, {Marcote}, {Mariotti},
  {Mart{\'\i}nez}, {Mazin}, {Menzel}, {Miranda}, {Mirzoyan}, {Moralejo},
  {Munar-Adrover}, {Nakajima}, {Niedzwiecki}, {Nilsson}, {Nishijima}, {Noda},
  {Orito}, {Overkemping}, {Paiano}, {Palatiello}, {Paneque}, {Paoletti},
  {Paredes}, {Paredes-Fortuny}, {Persic}, {Poutanen}, {Moroni}, {Prandini},
  {Puljak}, {Reinthal}, {Rhode}, {Rib{\'o}}, {Rico}, {Garcia}, {R{\"u}gamer},
  {Saito}, {Saito}, {Satalecka}, {Scalzotto}, {Scapin}, {Schultz}, {Schweizer},
  {Shore}, {Sillanp{\"a}{\"a}}, {Sitarek}, {Snidaric}, {Sobczynska}, {Spanier},
  {Stamatescu}, {Stamerra}, {Steinbring}, {Storz}, {Strzys}, {Takalo},
  {Takami}, {Tavecchio}, {Temnikov}, {Terzi{\'c}}, {Tescaro}, {Teshima},
  {Thaele}, {Tibolla}, {Torres}, {Toyama}, {Treves}, {Uellenbeck}, {Vogler},
  {Zanin}, {Kadler}, {Schulz}, {Ros}, {Bach}, {Krau{\ss}}, \&
  {Wilms}}]{Aleksic2014S}
{Aleksi{\'c}}, J., {Ansoldi}, S., {Antonelli}, L.~A., {et~al.} 2014, Science,
  346, 1080

\bibitem[{{Aliu} {et~al.}(2012){Aliu}, {Arlen}, {Aune}, {Beilicke}, {Benbow},
  {Bouvier}, {Bradbury}, {Buckley}, {Bugaev}, {Byrum}, {Cannon}, {Cesarini},
  {Ciupik}, {Collins-Hughes}, {Connolly}, {Cui}, {Dickherber}, {Duke},
  {Errando}, {Falcone}, {Finley}, {Finnegan}, {Fortson}, {Furniss}, {Galante},
  {Gall}, {Godambe}, {Griffin}, {Grube}, {Guenette}, {Gyuk}, {Hanna}, {Holder},
  {Huan}, {Hughes}, {Hui}, {Humensky}, {Imran}, {Kaaret}, {Karlsson},
  {Kertzman}, {Kieda}, {Krawczynski}, {Krennrich}, {Lang}, {LeBohec},
  {Madhavan}, {Maier}, {Majumdar}, {McArthur}, {McCann}, {Moriarty},
  {Mukherjee}, {Nu{\~n}ez}, {Ong}, {Orr}, {Otte}, {Park}, {Perkins}, {Pichel},
  {Pohl}, {Prokoph}, {Quinn}, {Ragan}, {Reyes}, {Reynolds}, {Roache}, {Rose},
  {Ruppel}, {Saxon}, {Schroedter}, {Sembroski}, {{\c{S}}ent{\"u}rk}, {Skole},
  {Staszak}, {Te{\v{s}}i{\'c}}, {Theiling}, {Thibadeau}, {Tsurusaki}, {Tyler},
  {Varlotta}, {Vassiliev}, {Vincent}, {Vivier}, {Wakely}, {Ward}, {Weekes},
  {Weinstein}, {Weisgarber}, {Williams}, \& {Zitzer}}]{veritas2012}
{Aliu}, E., {Arlen}, T., {Aune}, T., {et~al.} 2012, \apj, 746, 141

\bibitem[{{Beloborodov}(2017)}]{Beloborodov2017}
{Beloborodov}, A.~M. 2017, \apj, 850, 141

\bibitem[{{Blanch}(2021)}]{Magic2021}
{Blanch}, O. 2021, The Astronomer's Telegram, 14483, 1

\bibitem[{{Blandford} \& {Znajek}(1977)}]{BZ1977}
{Blandford}, R.~D., \& {Znajek}, R.~L. 1977, \mnras, 179, 433

\bibitem[{{Blumenthal} \& {Gould}(1970)}]{1970RvMP...42..237B}
{Blumenthal}, G.~R., \& {Gould}, R.~J. 1970, Reviews of Modern Physics, 42, 237

\bibitem[{Bransgrove {et~al.}(2021)Bransgrove, Ripperda, \&
  Philippov}]{Bransgrove2021}
Bransgrove, A., Ripperda, B., \& Philippov, A. 2021, Phys. Rev. Lett., 127,
  055101

\bibitem[{{Broderick} \& {Tchekhovskoy}(2015)}]{Broderick2015}
{Broderick}, A.~E., \& {Tchekhovskoy}, A. 2015, \apj, 809, 97

\bibitem[{{Cerutti} {et~al.}(2016){Cerutti}, {Philippov}, \&
  {Spitkovsky}}]{2016MNRAS.457.2401C}
{Cerutti}, B., {Philippov}, A.~A., \& {Spitkovsky}, A. 2016, \mnras, 457, 2401,
  \dodoi{10.1093/mnras/stw124}

\bibitem[{{Chashkina} {et~al.}(2021){Chashkina}, {Bromberg}, \&
  {Levinson}}]{Chashkina2021}
{Chashkina}, A., {Bromberg}, O., \& {Levinson}, A. 2021, \mnras

\bibitem[{{Chen} \& {Yuan}(2020)}]{chen2019}
{Chen}, A.~Y., \& {Yuan}, Y. 2020, \apj, 895, 121

\bibitem[{{Chernoglazov} {et~al.}(in prep.){Chernoglazov}, {Hakobyan}, \&
  {Philippov}}]{chernoglazovinprep}
{Chernoglazov}, A., {Hakobyan}, H., \& {Philippov}, A. in prep.

\bibitem[{{Crinquand} {et~al.}(2020){Crinquand}, {Cerutti}, {Philippov},
  {Parfrey}, \& {Dubus}}]{Crinquand2020}
{Crinquand}, B., {Cerutti}, B., {Philippov}, A., {Parfrey}, K., \& {Dubus}, G.
  2020, \prl, 124, 145101

\bibitem[{{EHT MWL Science Working Group} {et~al.}(2021){EHT MWL Science
  Working Group}, {Algaba}, {Anczarski}, {Asada}, {Balokovi{\'c}}, {Chandra},
  {Cui}, {Falcone}, {Giroletti}, {Goddi}, {Hada}, {Haggard}, {Jorstad}, {Kaur},
  {Kawashima}, {Keating}, {Kim}, {Kino}, {Komossa}, {Kravchenko}, {Krichbaum},
  {Lee}, {Lu}, {Lucchini}, {Markoff}, {Neilsen}, {Nowak}, {Park}, {Principe},
  {Ramakrishnan}, {Reynolds}, {Sasada}, {Savchenko}, {Williamson}, {Event
  Horizon Telescope Collaboration}, {Akiyama}, {Alberdi}, {Alef}, {Anantua},
  {Azulay}, {Baczko}, {Ball}, {Barrett}, {Bintley}, {Benson}, {Blackburn},
  {Blundell}, {Boland}, {Bouman}, {Bower}, {Boyce}, {Bremer}, {Brinkerink},
  {Brissenden}, {Britzen}, {Broderick}, {Broguiere}, {Bronzwaer}, {Byun},
  {Carlstrom}, {Chael}, {Chan}, {Chatterjee}, {Chatterjee}, {Chen}, {Chen},
  {Chesler}, {Cho}, {Christian}, {Conway}, {Cordes}, {Crawford}, {Crew},
  {Cruz-Osorio}, {Davelaar}, {de Laurentis}, {Deane}, {Dempsey}, {Desvignes},
  {Dexter}, {Doeleman}, {Eatough}, {Falcke}, {Farah}, {Fish}, {Fomalont},
  {Ford}, {Fraga-Encinas}, {Friberg}, {Fromm}, {Fuentes}, {Galison}, {Gammie},
  {Garc{\'\i}a}, {Gentaz}, {Georgiev}, {Gold}, {G{\'o}mez}, {G{\'o}mez-Ruiz},
  {Gu}, {Gurwell}, {Hecht}, {Hesper}, {Ho}, {Ho}, {Honma}, {Huang}, {Huang},
  {Hughes}, {Ikeda}, {Inoue}, {Issaoun}, {James}, {Jannuzi}, {Janssen},
  {Jeter}, {Jiang}, {Jim{\'e}nez-Rosales}, {Johnson}, {Jung}, {Karami},
  {Karuppusamy}, {Kettenis}, {Kim}, {Kim}, {Kim}, {Koay}, {Kofuji}, {Koch},
  {Koyama}, {Kramer}, {Kramer}, {Kuo}, {Lauer}, {Levis}, {Li}, {Li},
  {Lindqvist}, {Lico}, {Lindahl}, {Liu}, {Liu}, {Liuzzo}, {Lo}, {Lobanov},
  {Loinard}, {Lonsdale}, {MacDonald}, {Mao}, {Marchili}, {Marrone}, {Marscher},
  {Mart{\'\i}-Vidal}, {Matsushita}, {Matthews}, {Medeiros}, {Menten}, {Mizuno},
  {Mizuno}, {Moran}, {Moriyama}, {Moscibrodzka}, {M{\"u}ller}, {Musoke},
  {Mej{\'\i}as}, {Nagai}, {Nagar}, {Nakamura}, {Narayan}, {Narayanan},
  {Natarajan}, {Nathanail}, {Neri}, {Ni}, {Noutsos}, {Okino}, {Olivares},
  {Ortiz-Le{\'o}n}, {Oyama}, {{\"O}zel}, {Palumbo}, {Patel}, {Pen}, {Pesce},
  {Pi{\'e}tu}, {Plambeck}, {Popstefanija}, {Porth}, {P{\"o}tzl}, {Prather},
  {Preciado-L{\'o}pez}, {Psaltis}, {Pu}, {Rao}, {Rawlings}, {Raymond},
  {Rezzolla}, {Ricarte}, {Ripperda}, {Roelofs}, {Rogers}, {Ros}, {Rose},
  {Roshanineshat}, {Rottmann}, {Roy}, {Ruszczyk}, {Rygl}, {S{\'a}nchez},
  {S{\'a}nchez-Arguelles}, {Savolainen}, {Schloerb}, {Schuster}, {Shao},
  {Shen}, {Small}, {Sohn}, {Soohoo}, {Sun}, {Tazaki}, {Tetarenko}, {Tiede},
  {Tilanus}, {Titus}, {Toma}, {Torne}, {Trent}, {Traianou}, {Trippe}, {van
  Bemmel}, {van Langevelde}, {van Rossum}, {Wagner}, {Ward-Thompson}, {Wardle},
  {Weintroub}, {Wex}, {Wharton}, {Wielgus}, {Wong}, {Wu}, {Yoon}, {Young},
  {Young}, {Younsi}, {Yuan}, {Yuan}, {Zensus}, {Zhao}, {Zhao}, {Fermi Large
  Area Telescope Collaboration}, {Principe}, {Giroletti}, {D'Ammando},
  {Orienti}, {H.~E.~S.~S. Collaboration}, {Abdalla}, {Adam}, {Aharonian},
  {Benkhali}, {Ang{\"u}ner}, {Arcaro}, {Armand}, {Armstrong}, {Ashkar},
  {Backes}, {Baghmanyan}, {Barbosa Martins}, {Barnacka}, {Barnard},
  {Becherini}, {Berge}, {Bernl{\"o}hr}, {Bi}, {B{\"o}ttcher}, {Boisson},
  {Bolmont}, {de Lavergne}, {Breuhaus}, {Brun}, {Brun}, {Bryan}, {B{\"u}chele},
  {Bulik}, {Bylund}, {Caroff}, {Carosi}, {Casanova}, {Chand}, {Chen}, {Cotter},
  {Cury{\l}o}, {Damascene Mbarubucyeye}, {Davids}, {Davies}, {Deil}, {Devin},
  {Dewilt}, {Dirson}, {Djannati-Ata{\"\i}}, {Dmytriiev}, {Donath},
  {Doroshenko}, {Duffy}, {Dyks}, {Egberts}, {Eichhorn}, {Einecke}, {Emery},
  {Ernenwein}, {Feijen}, {Fegan}, {Fiasson}, {de Clairfontaine}, {Fontaine},
  {Funk}, {F{\"u}{\ss}ling}, {Gabici}, {Gallant}, {Giavitto}, {Giunti},
  {Glawion}, {Glicenstein}, {Gottschall}, {Grondin}, {Hahn}, {Haupt},
  {Hermann}, {Hinton}, {Hofmann}, {Hoischen}, {Holch}, {Holler}, {H{\"o}rbe},
  {Horns}, {Huber}, {Jamrozy}, {Jankowsky}, {Jankowsky}, {Jardin-Blicq},
  {Joshi}, {Jung-Richardt}, {Kasai}, {Kastendieck}, {Katarzy{\'n}ski}, {Katz},
  {Khangulyan}, {Kh{\'e}lifi}, {Klepser}, {Klu{\'z}niak}, {Komin}, {Konno},
  {Kosack}, {Kostunin}, {Kreter}, {Lamanna}, {Lemi{\`e}re}, {Lemoine-Goumard},
  {Lenain}, {Levy}, {Lohse}, {Lypova}, {Mackey}, {Majumdar}, {Malyshev},
  {Malyshev}, {Marandon}, {Marchegiani}, {Marcowith}, {Mares},
  {Mart{\'\i}-Devesa}, {Marx}, {Maurin}, {Meintjes}, {Meyer}, {Moderski},
  {Mohamed}, {Mohrmann}, {Montanari}, {Moore}, {Morris}, {Moulin}, {Muller},
  {Murach}, {Nakashima}, {Nayerhoda}, {de Naurois}, {Ndiyavala},
  {Niederwanger}, {Niemiec}, {Oakes}, {O'Brien}, {Odaka}, {Ohm},
  {Olivera-Nieto}, {de Ona Wilhelmi}, {Ostrowski}, {Panter}, {Panny},
  {Parsons}, {Peron}, {Peyaud}, {Piel}, {Pita}, {Poireau}, {Noel}, {Prokhorov},
  {Prokoph}, {P{\"u}hlhofer}, {Punch}, {Quirrenbach}, {Rauth}, {Reichherzer},
  {Reimer}, {Reimer}, {Remy}, {Renaud}, {Rieger}, {Rinchiuso}, {Romoli},
  {Rowell}, {Rudak}, {Ruiz-Velasco}, {Sahakian}, {Sailer}, {Sanchez},
  {Santangelo}, {Sasaki}, {Scalici}, {Schutte}, {Schwanke}, {Schwemmer},
  {Seglar-Arroyo}, {Senniappan}, {Seyffert}, {Shafi}, {Shiningayamwe},
  {Simoni}, {Sinha}, {Sol}, {Specovius}, {Spencer}, {Spir-Jacob}, {Stawarz},
  {Sun}, {Steenkamp}, {Stegmann}, {Steinmassl}, {Steppa}, {Takahashi},
  {Tavernier}, {Taylor}, {Terrier}, {Tiziani}, {Tluczykont}, {Tomankova},
  {Trichard}, {Tsirou}, {Tuffs}, {Uchiyama}, {van der Walt}, {van Eldik}, {van
  Rensburg}, {van Soelen}, {Vasileiadis}, {Veh}, {Venter}, {Vincent}, {Vink},
  {V{\"o}lk}, {Vuillaume}, {Wadiasingh}, {Wagner}, {Watson}, {Werner}, {White},
  {Wierzcholska}, {Wong}, {Yusafzai}, {Zacharias}, {Zanin}, {Zargaryan},
  {Zdziarski}, {Zech}, {Zhu}, {Zorn}, {Zouari}, {{\.Z}ywucka}, {MAGIC
  Collaboration}, {Acciari}, {Ansoldi}, {Antonelli}, {Engels}, {Artero},
  {Asano}, {Baack}, {Babi{\'c}}, {Baquero}, {de Almeida}, {Barrio}, {Becerra
  Gonz{\'a}lez}, {Bednarek}, {Bellizzi}, {Bernardini}, {Bernardos}, {Berti},
  {Besenrieder}, {Bhattacharyya}, {Bigongiari}, {Biland}, {Blanch}, {Bonnoli},
  {Bo{\v{s}}njak}, {Busetto}, {Carosi}, {Ceribella}, {Cerruti}, {Chai},
  {Chilingarian}, {Cikota}, {Colak}, {Colombo}, {Contreras}, {Cortina},
  {Covino}, {D'Amico}, {D'Elia}, {da Vela}, {Dazzi}, {de Angelis}, {de Lotto},
  {Delfino}, {Delgado}, {Delgado Mendez}, {Depaoli}, {di Pierro}, {di Venere},
  {Do Souto Espi{\~n}eira}, {Dominis Prester}, {Donini}, {Dorner}, {Doro},
  {Elsaesser}, {Ramazani}, {Fattorini}, {Ferrara}, {Fonseca}, {Font}, {Fruck},
  {Fukami}, {Garc{\'\i}a L{\'o}pez}, {Garczarczyk}, {Gasparyan}, {Gaug},
  {Giglietto}, {Giordano}, {Gliwny}, {Godinovi{\'c}}, {Green}, {Green},
  {Hadasch}, {Hahn}, {Heckmann}, {Herrera}, {Hoang}, {Hrupec}, {H{\"u}tten},
  {Inada}, {Inoue}, {Ishio}, {Iwamura}, {Jim{\'e}nez}, {Jormanainen}, {Jouvin},
  {Kajiwara}, {Karjalainen}, {Kerszberg}, {Kobayashi}, {Kubo}, {Kushida},
  {Lamastra}, {Lelas}, {Leone}, {Lindfors}, {Lombardi}, {Longo},
  {L{\'o}pez-Coto}, {L{\'o}pez-Moya}, {L{\'o}pez-Oramas}, {Loporchio}, {Machado
  de Oliveira Fraga}, {Maggio}, {Majumdar}, {Makariev}, {Mallamaci}, {Maneva},
  {Manganaro}, {Mannheim}, {Maraschi}, {Mariotti}, {Mart{\'\i}nez}, {Mazin},
  {Menchiari}, {Mender}, {Mi{\'c}anovi{\'c}}, {Miceli}, {Miener}, {Minev},
  {Miranda}, {Mirzoyan}, {Molina}, {Moralejo}, {Morcuende}, {Moreno},
  {Moretti}, {Neustroev}, {Nigro}, {Nilsson}, {Nishijima}, {Noda}, {Nozaki},
  {Ohtani}, {Oka}, {Otero-Santos}, {Paiano}, {Palatiello}, {Paneque},
  {Paoletti}, {Paredes}, {Pavleti{\'c}}, {Pe{\~n}il}, {Perennes}, {Persic},
  {Moroni}, {Prandini}, {Priyadarshi}, {Puljak}, {Rhode}, {Rib{\'o}}, {Rico},
  {Righi}, {Rugliancich}, {Saha}, {Sahakyan}, {Saito}, {Sakurai}, {Satalecka},
  {Saturni}, {Schleicher}, {Schmidt}, {Schweizer}, {Sitarek},
  {{\v{S}}nidari{\'c}}, {Sobczynska}, {Spolon}, {Stamerra}, {Strom}, {Strzys},
  {Suda}, {Suri{\'c}}, {Takahashi}, {Tavecchio}, {Temnikov}, {Terzi{\'c}},
  {Teshima}, {Tosti}, {Truzzi}, {Tutone}, {Ubach}, {van Scherpenberg}, {Vanzo},
  {Vazquez Acosta}, {Ventura}, {Verguilov}, {Vigorito}, {Vitale}, {Vovk},
  {Will}, {Wunderlich}, {Zari{\'c}}, {VERITAS Collaboration}, {Adams},
  {Benbow}, {Brill}, {Capasso}, {Christiansen}, {Chromey}, {Daniel}, {Errando},
  {Farrell}, {Feng}, {Finley}, {Fortson}, {Furniss}, {Gent}, {Giuri}, {Hassan},
  {Hervet}, {Holder}, {Hughes}, {Humensky}, {Jin}, {Kaaret}, {Kertzman},
  {Kieda}, {Kumar}, {Lang}, {Lundy}, {Maier}, {Moriarty}, {Mukherjee}, {Nieto},
  {Nievas-Rosillo}, {O'Brien}, {Ong}, {Otte}, {Patel}, {Pfrang}, {Pohl},
  {Prado}, {Pueschel}, {Quinn}, {Ragan}, {Reynolds}, {Ribeiro}, {Richards},
  {Roache}, {Rulten}, {Ryan}, {Santander}, {Sembroski}, {Shang}, {Weinstein},
  {Williams}, {Williamson}, {Eavn Collaboration}, {Hirota}, {Cui}, {Niinuma},
  {Ro}, {Sakai}, {Sawada-Satoh}, {Wajima}, {Wang}, {Liu}, \&
  {Yonekura}}]{MWL2021}
{EHT MWL Science Working Group}, {Algaba}, J.~C., {Anczarski}, J., {et~al.}
  2021, \apjl, 911, L11

\bibitem[{{Event Horizon Telescope Collaboration} {et~al.}(2021){Event Horizon
  Telescope Collaboration}, {Akiyama}, {Algaba}, {Alberdi}, {Alef}, {Anantua},
  {Asada}, {Azulay}, {Baczko}, {Ball}, {Balokovi{\'c}}, {Barrett}, {Benson},
  {Bintley}, {Blackburn}, {Blundell}, {Boland}, {Bouman}, {Bower}, {Boyce},
  {Bremer}, {Brinkerink}, {Brissenden}, {Britzen}, {Broderick}, {Broguiere},
  {Bronzwaer}, {Byun}, {Carlstrom}, {Chael}, {Chan}, {Chatterjee},
  {Chatterjee}, {Chen}, {Chen}, {Chesler}, {Cho}, {Christian}, {Conway},
  {Cordes}, {Crawford}, {Crew}, {Cruz-Osorio}, {Cui}, {Davelaar}, {De
  Laurentis}, {Deane}, {Dempsey}, {Desvignes}, {Dexter}, {Doeleman}, {Eatough},
  {Falcke}, {Farah}, {Fish}, {Fomalont}, {Ford}, {Fraga-Encinas}, {Friberg},
  {Fromm}, {Fuentes}, {Galison}, {Gammie}, {Garc{\'\i}a}, {Gelles}, {Gentaz},
  {Georgiev}, {Goddi}, {Gold}, {G{\'o}mez}, {G{\'o}mez-Ruiz}, {Gu}, {Gurwell},
  {Hada}, {Haggard}, {Hecht}, {Hesper}, {Himwich}, {Ho}, {Ho}, {Honma},
  {Huang}, {Huang}, {Hughes}, {Ikeda}, {Inoue}, {Issaoun}, {James}, {Jannuzi},
  {Janssen}, {Jeter}, {Jiang}, {Jimenez-Rosales}, {Johnson}, {Jorstad}, {Jung},
  {Karami}, {Karuppusamy}, {Kawashima}, {Keating}, {Kettenis}, {Kim}, {Kim},
  {Kim}, {Kim}, {Kino}, {Koay}, {Kofuji}, {Koch}, {Koyama}, {Kramer}, {Kramer},
  {Krichbaum}, {Kuo}, {Lauer}, {Lee}, {Levis}, {Li}, {Li}, {Lindqvist}, {Lico},
  {Lindahl}, {Liu}, {Liu}, {Liuzzo}, {Lo}, {Lobanov}, {Loinard}, {Lonsdale},
  {Lu}, {MacDonald}, {Mao}, {Marchili}, {Markoff}, {Marrone}, {Marscher},
  {Mart{\'\i}-Vidal}, {Matsushita}, {Matthews}, {Medeiros}, {Menten}, {Mizuno},
  {Mizuno}, {Moran}, {Moriyama}, {Moscibrodzka}, {M{\"u}ller}, {Musoke}, {Mus
  Mej{\'\i}as}, {Michalik}, {Nadolski}, {Nagai}, {Nagar}, {Nakamura},
  {Narayan}, {Narayanan}, {Natarajan}, {Nathanail}, {Neilsen}, {Neri}, {Ni},
  {Noutsos}, {Nowak}, {Okino}, {Olivares}, {Ortiz-Le{\'o}n}, {Oyama},
  {{\"O}zel}, {Palumbo}, {Park}, {Patel}, {Pen}, {Pesce}, {Pi{\'e}tu},
  {Plambeck}, {PopStefanija}, {Porth}, {P{\"o}tzl}, {Prather},
  {Preciado-L{\'o}pez}, {Psaltis}, {Pu}, {Ramakrishnan}, {Rao}, {Rawlings},
  {Raymond}, {Rezzolla}, {Ricarte}, {Ripperda}, {Roelofs}, {Rogers}, {Ros},
  {Rose}, {Roshanineshat}, {Rottmann}, {Roy}, {Ruszczyk}, {Rygl},
  {S{\'a}nchez}, {S{\'a}nchez-Arguelles}, {Sasada}, {Savolainen}, {Schloerb},
  {Schuster}, {Shao}, {Shen}, {Small}, {Sohn}, {SooHoo}, {Sun}, {Tazaki},
  {Tetarenko}, {Tiede}, {Tilanus}, {Titus}, {Toma}, {Torne}, {Trent},
  {Traianou}, {Trippe}, {van Bemmel}, {van Langevelde}, {van Rossum}, {Wagner},
  {Ward-Thompson}, {Wardle}, {Weintroub}, {Wex}, {Wharton}, {Wielgus}, {Wong},
  {Wu}, {Yoon}, {Young}, {Young}, {Younsi}, {Yuan}, {Yuan}, {Zensus}, {Zhao},
  \& {Zhao}}]{EHTVII2021}
{Event Horizon Telescope Collaboration}, {Akiyama}, K., {Algaba}, J.~C.,
  {et~al.} 2021, \apjl, 910, L13

\bibitem[{{Guo} {et~al.}(2014){Guo}, {Li}, {Daughton}, \&
  {Liu}}]{2014PhRvL.113o5005G}
{Guo}, F., {Li}, H., {Daughton}, W., \& {Liu}, Y.-H. 2014, \prl, 113, 155005

\bibitem[{{Guo} {et~al.}(2015){Guo}, {Liu}, {Daughton}, \&
  {Li}}]{2015ApJ...806..167G}
{Guo}, F., {Liu}, Y.-H., {Daughton}, W., \& {Li}, H. 2015, \apj, 806, 167

\bibitem[{{Hakobyan} {et~al.}(2021){Hakobyan}, {Petropoulou}, {Spitkovsky}, \&
  {Sironi}}]{2021ApJ...912...48H}
{Hakobyan}, H., {Petropoulou}, M., {Spitkovsky}, A., \& {Sironi}, L. 2021,
  \apj, 912, 48

\bibitem[{{Hakobyan} {et~al.}(2019{\natexlab{a}}){Hakobyan}, {Philippov}, \&
  {Spitkovsky}}]{2019ApJ...877...53H}
{Hakobyan}, H., {Philippov}, A., \& {Spitkovsky}, A. 2019{\natexlab{a}}, \apj,
  877, 53

\bibitem[{{Hakobyan} {et~al.}(2019{\natexlab{b}}){Hakobyan}, {Philippov}, \&
  {Spitkovsky}}]{Hakobyan2019}
---. 2019{\natexlab{b}}, \apj, 877, 53

\bibitem[{Hakobyan \& Spitkovsky(2020)}]{tristanv2}
Hakobyan, H., \& Spitkovsky, A. 2020, Tristan-MP v2, multi-species
  particle-in-cell plasma code:
  \texttt{github.com/PrincetonUniversity/tristan-mp-v2}.
\newblock \url{https://princetonuniversity.github.io/tristan-v2/}

\bibitem[{{Harris} {et~al.}(2011){Harris}, {Massaro}, {Cheung}, {Horns},
  {Raue}, {Stawarz}, {Wagner}, {Colin}, {Mazin}, {Wagner}, {Beilicke},
  {LeBohec}, {Hui}, \& {Mukherjee}}]{2011ApJ...743..177H}
{Harris}, D.~E., {Massaro}, F., {Cheung}, C.~C., {et~al.} 2011, \apj, 743, 177

\bibitem[{Harris(1962)}]{Harris1962}
Harris, E.~G. 1962, Il Nuovo Cimento (1955-1965), 23, 115

\bibitem[{{Jia} \& {et al.}(in prep.)}]{jiainprep}
{Jia}, H., \& {et al.} in prep.

\bibitem[{{Kisaka} {et~al.}(2022){Kisaka}, {Levinson}, {Toma}, \&
  {Niv}}]{Kisaka22}
{Kisaka}, S., {Levinson}, A., {Toma}, K., \& {Niv}, I. 2022, \apj, 924, 28

\bibitem[{{Landau} \& {Lifshitz}(1975)}]{1975ctf..book.....L}
{Landau}, L.~D., \& {Lifshitz}, E.~M. 1975, {The classical theory of fields}

\bibitem[{{Lefa} {et~al.}(2012){Lefa}, {Kelner}, \&
  {Aharonian}}]{2012ApJ...753..176L}
{Lefa}, E., {Kelner}, S.~R., \& {Aharonian}, F.~A. 2012, \apj, 753, 176

\bibitem[{{Levinson} \& {Rieger}(2011)}]{Levinson2011}
{Levinson}, A., \& {Rieger}, F. 2011, \apj, 730, 123

\bibitem[{{Lyubarskii}(1996)}]{1996A&A...311..172L}
{Lyubarskii}, Y.~E. 1996, \aap, 311, 172

\bibitem[{{Moscibrodzka} {et~al.}(2011){Moscibrodzka}, {Gammie}, {Dolence}, \&
  {Shiokawa}}]{Moscibrodzka2011}
{Moscibrodzka}, M., {Gammie}, C.~F., {Dolence}, J.~C., \& {Shiokawa}, H. 2011,
  \apj, 735, 9

\bibitem[{Narayan {et~al.}(2003)Narayan, Igumenshchev, \&
  Abramowicz}]{narayan2003}
Narayan, R., Igumenshchev, I.~V., \& Abramowicz, M.~A. 2003, Publications of
  the Astronomical Society of Japan, 55, L69

\bibitem[{{Petropoulou} \& {Sironi}(2018)}]{2018MNRAS.481.5687P}
{Petropoulou}, M., \& {Sironi}, L. 2018, \mnras, 481, 5687

\bibitem[{{Prieto} {et~al.}(2016){Prieto}, {Fern{\'a}ndez-Ontiveros},
  {Markoff}, {Espada}, \& {Gonz{\'a}lez-Mart{\'\i}n}}]{Prieto2016}
{Prieto}, M.~A., {Fern{\'a}ndez-Ontiveros}, J.~A., {Markoff}, S., {Espada}, D.,
  \& {Gonz{\'a}lez-Mart{\'\i}n}, O. 2016, \mnras, 457, 3801

\bibitem[{{Ripperda} {et~al.}(2020){Ripperda}, {Bacchini}, \&
  {Philippov}}]{Ripperda2020}
{Ripperda}, B., {Bacchini}, F., \& {Philippov}, A.~A. 2020, \apj, 900, 100

\bibitem[{{Ripperda} {et~al.}(2022){Ripperda}, {Liska}, {Chatterjee}, {Musoke},
  {Philippov}, {Markoff}, {Tchekhovskoy}, \& {Younsi}}]{Ripperda2022}
{Ripperda}, B., {Liska}, M., {Chatterjee}, K., {et~al.} 2022, \apjl, 924, L32

\bibitem[{{Ripperda} {et~al.}(2019){Ripperda}, {Porth}, {Sironi}, \&
  {Keppens}}]{ripperda2019}
{Ripperda}, B., {Porth}, O., {Sironi}, L., \& {Keppens}, R. 2019, \mnras, 485,
  299

\bibitem[{{Rybicki} \& {Lightman}(1979)}]{1979rpa..book.....R}
{Rybicki}, G.~B., \& {Lightman}, A.~P. 1979, {Radiative processes in
  astrophysics}

\bibitem[{{Sironi} \& {Beloborodov}(2020)}]{2020ApJ...899...52S}
{Sironi}, L., \& {Beloborodov}, A.~M. 2020, \apj, 899, 52

\bibitem[{Sironi {et~al.}(2016)Sironi, Giannios, \& Petropoulou}]{sironi2016}
Sironi, L., Giannios, D., \& Petropoulou, M. 2016, MNRAS, 462, 1

\bibitem[{{Sironi} \& {Spitkovsky}(2014)}]{2014ApJ...783L..21S}
{Sironi}, L., \& {Spitkovsky}, A. 2014, \apjl, 783, L21

\bibitem[{Sironi \& Spitkovsky(2014)}]{sironi2014}
Sironi, L., \& Spitkovsky, A. 2014, ApJL, 783, 1

\bibitem[{Stanzione {et~al.}(2020)Stanzione, West, Evans, Minyard, Ghattas, \&
  Panda}]{Frontera}
Stanzione, D., West, J., Evans, R.~T., {et~al.} 2020, in Practice and
  Experience in Advanced Research Computing, PEARC '20 (New York, NY, USA:
  Association for Computing Machinery), 106–111

\bibitem[{{Sun} {et~al.}(2018){Sun}, {Yang}, {Rieger}, {Liu}, \&
  {Aharonian}}]{2018A&A...612A.106S}
{Sun}, X.-N., {Yang}, R.-Z., {Rieger}, F.~M., {Liu}, R.-Y., \& {Aharonian}, F.
  2018, \aap, 612, A106

\bibitem[{{Svensson}(1987)}]{1987MNRAS.227..403S}
{Svensson}, R. 1987, \mnras, 227, 403, \dodoi{10.1093/mnras/227.2.403}

\bibitem[{{Tchekhovskoy} {et~al.}(2011){Tchekhovskoy}, {Narayan}, \&
  {McKinney}}]{Tchekhovskoy2011}
{Tchekhovskoy}, A., {Narayan}, R., \& {McKinney}, J.~C. 2011, \mnras, 418, L79

\bibitem[{{Uzdensky} \& {Spitkovsky}(2014)}]{2014ApJ...780....3U}
{Uzdensky}, D.~A., \& {Spitkovsky}, A. 2014, \apj, 780, 3

\bibitem[{{Werner} {et~al.}(2019){Werner}, {Philippov}, \&
  {Uzdensky}}]{2019MNRAS.482L..60W}
{Werner}, G.~R., {Philippov}, A.~A., \& {Uzdensky}, D.~A. 2019, \mnras, 482,
  L60

\bibitem[{{Werner} {et~al.}(2016){Werner}, {Uzdensky}, {Cerutti}, {Nalewajko},
  \& {Begelman}}]{2016ApJ...816L...8W}
{Werner}, G.~R., {Uzdensky}, D.~A., {Cerutti}, B., {Nalewajko}, K., \&
  {Begelman}, M.~C. 2016, \apjl, 816, L8

\bibitem[{{Wong} {et~al.}(2021){Wong}, {Ryan}, \&
  {Gammie}}]{2021ApJ...907...73W}
{Wong}, G.~N., {Ryan}, B.~R., \& {Gammie}, C.~F. 2021, \apj, 907, 73

\bibitem[{{Yao} {et~al.}(2021){Yao}, {Dexter}, {Chen}, {Ryan}, \&
  {Wong}}]{2021MNRAS.507.4864Y}
{Yao}, P.~Z., {Dexter}, J., {Chen}, A.~Y., {Ryan}, B.~R., \& {Wong}, G.~N.
  2021, \mnras, 507, 4864

\bibitem[{{Zhang} {et~al.}(2021){Zhang}, {Sironi}, \&
  {Giannios}}]{2021ApJ...922..261Z}
{Zhang}, H., {Sironi}, L., \& {Giannios}, D. 2021, \apj, 922, 261

\bibitem[{{Zirakashvili} \& {Aharonian}(2007)}]{2007A&A...465..695Z}
{Zirakashvili}, V.~N., \& {Aharonian}, F. 2007, \aap, 465, 695,
  \dodoi{10.1051/0004-6361:20066494}

\end{thebibliography}
\bibliographystyle{aasjournal}

\clearpage

\end{document}